\documentclass[aps,prd,twocolumn,preprintnumbers,nofootinbib,superscriptaddress,amsmath]{revtex4-2}
\usepackage[english]{babel}
\usepackage{bm}
\usepackage[utf8]{inputenc}
\usepackage[colorinlistoftodos, color=green!40, prependcaption]{todonotes}
\usepackage{amssymb}
\usepackage{graphicx}
\usepackage{natbib}
\usepackage[normalem]{ulem}

\usepackage[pdftex, pdftitle={Article}, pdfauthor={Author}]{hyperref}

\begin{document}

\title{ACT DR6+Planck impact on inflation with non-zero vacuum expectation values and the post-inflationary behavior}

\author{F. B. M. dos Santos}\email{fbmsantos@on.br}
\affiliation{Observatório Nacional, Rio de Janeiro - RJ, 20921-400, Brasil}
\author{J. G. Rodrigues}\email{jamersonrodrigues@on.br}
\affiliation{Observatório Nacional, Rio de Janeiro - RJ, 20921-400, Brasil}
\affiliation{Departamento de Física Teórica e Experimental, Universidade Federal do Rio Grande do Norte, Natal - RN, 59072-970, Brasil}
\author{G. Rodrigues}\email{gabrielrodrigues@on.br}
\affiliation{Observatório Nacional, Rio de Janeiro - RJ, 20921-400, Brasil}
\author{C. Siqueira}\email{csiqueira@on.br}
\affiliation{Observatório Nacional, Rio de Janeiro - RJ, 20921-400, Brasil}
\author{J. S. Alcaniz}\email{alcaniz@on.br}
\affiliation{Observatório Nacional, Rio de Janeiro - RJ, 20921-400, Brasil}

\begin{abstract}
    The impact of the most recent cosmic microwave background (CMB) data from the Atacama Cosmology Telescope (ACT) is studied for a model of cosmic inflation which predicts a non-zero vacuum expectation value (VEV) $M$ for a large-field regime. Since lower values of $M$ are compatible with the higher spectral index $n_s$ provided by the ACT+Planck joint analysis, we establish new limits on this parameter while also considering further CMB data from the latest BICEP/Keck Array release for CMB polarization modes. We find $\log_{10}M/M_{Pl}=-2.5^{+1.1}_{-1.3}$ at 68\% confidence level, compatible with $M/M_{Pl}\simeq 0.003$, which is interesting for post-inflationary processes, such as preheating. We conduct a lattice simulation for the inflaton field for the first few e-folds, as the model is compatible with the production of relics such as oscillons, which are possible candidates as sources of gravitational waves and primordial black holes. We find that the model indeed produces localized, quasi-spherical structures compatible with oscillons, which might lead to signatures detectable by future experiments. However, in agreement with recent works, we find that although the abundance of gravitational waves that could be generated in this regime has an amplitude well within the sensitivities of these detectors, the frequency range is on the GHz limit, away from the expected frequencies. Finally, we estimate the impact of a coupling of the type $y\phi\chi^2$ to the inflaton, in the realization of perturbative reheating, directly impacting the predictions of the model, as lower values of $M$ are consistent with both the entire allowed temperature range, and the limits imposed by BICEP/Keck Array+Planck+ACT.
\end{abstract}

\maketitle

\section{Introduction}

The inflationary period, a phase of accelerating expansion that sets proper initial conditions in the early universe \cite{Starobinsky:1980te,Guth:1980zm,Linde:1981mu}, is connected with the production of particles of the Standard Model (SM) through the  reheating \cite{Albrecht:1982mp,Abbott:1982hn,Kofman:1994rk,Kofman:1997yn}. In the canonical conjecture, the transition to a radiation-dominated phase is described following the perturbative decay of the inflaton field, with possible decay channels determined, as usual, by the internal symmetries of the system. On the other hand, it has been found that the non-perturbative behavior of fields immediately after inflation can play a significant role in particle production, including post-inflationary relics potentially detectable by future experiments, such as oscillons, non-topological objects produced by scalar fields \cite{Gleiser:1993pt,Copeland:1995fq}.

Oscillons can be produced in the cosmological regime after inflation, where the field starts oscillating around the minimum of its potential $V(\phi)$. However, there are some conditions that must be fulfilled in order to have the presence of these relics; the first condition is that the potential must approximate a quadratic form around its minimum, but shallower than this quadratic form for larger values of $\phi$. Additionally, it is important that the potential supports the tachyonic resonance that leads to the usual preheating process, in which the self-resonant inflaton has its perturbations amplified by oscillations, consequently leading to a fragmentation of the scalar field, where the coherent oscillatory regime is broken \cite{Amin:2010jq,Amin:2011hj,Amin:2014eta,Antusch:2016con,Lozanov:2017hjm,Cotner:2018vug,Antusch:2019qrr,Piani:2023aof,Mahbub:2023faw,Kasai:2025coe}. While the preheating process can lead to a quick radiation dominance, the fact that the minimum of $V(\phi)$ is nearly quadratic during oscillon formation means that these objects behave approximately as matter ($w\simeq 0$) \cite{Lozanov:2017hjm}. This can lead to different reheating periods, consequently affecting the posterior evolution of the Universe. Also, due to their quasi-spherical shape, oscillons can be regarded as a potential source of gravitational waves from the early Universe \cite{Zhou:2013tsa,Liu:2017hua,Antusch:2017vga,Lozanov:2019ylm,Hiramatsu:2020obh,Lozanov:2022yoy,Drees:2025iue}.

The relationship between reheating predictions and the latest CMB data is a current topic of investigation, especially following the recent release from the Atacama Cosmology Telescope (ACT) \cite{Louis_2025}, which maps the background radiation for temperature and polarization on smaller scales. The reason is that one of the main predictions of inflation, given by a slowly-rolling scalar field, is that the primordial power spectrum, while almost invariant with the scale, has the Harrison-Zeldovich scale-invariant spectrum case ($n_s = 1$) heavily excluded by the current Planck data, at more than $8\sigma$ confidence level (C.L.). However, the new ACT data leads to an intriguing result that, while still compatible with the non-scale invariance, leads to a higher value of $n_s=0.974\pm 0.003$, for a combined analysis with Planck \cite{Planck:2018jri} and baryonic acoustic oscillation (BAO) Dark Energy Spectroscopic Instrument (DESI) data \cite{DESI:2024mwx,DESI:2025zgx}. Naturally, this led to a series of reevaluations of cosmic inflationary models \cite{Zharov:2025zjg,lq71:b84v,Haque:2025uri,Peng:2025bws,Addazi:2025qra,Berera:2025vsu,Gao:2025onc,Yogesh:2025wak,Liu:2025qca,Drees:2025ngb,Kallosh:2025rni}, since in light of these results, even established scenarios such as the Starobinsky inflation would be disfavored by the new data.

In this work, we investigate the impact of Planck+ACT-DR6 data, the formation of oscillons, and their consequences for gravitational wave emission.
As a representative model, we investigate a simple potential, motivated by supersymmetry, the Witten-O'Raifeartaigh (WR) model \cite{ORaifeartaigh:1975nky,Witten:1981kv,Albrecht:1983ib,Martin:2013tda}. Its dependence on a mass scale that defines the position of its minimum allows us to analyze the impact of this choice when considering current observational constraints and the limits imposed by the reheating process. We will see that, interestingly, the model's predictions are compatible with the current constraints on inflation from the combined analysis of Planck+ACT-DR6, which favors a flatter primordial power spectrum, as indicated by a higher spectral index. Such results lead to a restored viability of the model given the data, without resorting to extensions such as the non-minimal coupling of the field to gravity, for instance. Moreover, the possibility of tachyonic amplification of the field after inflation motivates the use of lattice simulations to model the behavior of the field and check for the presence of post-inflationary relics, using the results from the statistical parameter inference performed.

This work is organized as follows. In Section \ref{sec:2}, we briefly discuss the WR model and its main prediction for inflation, followed by the methodology and results of the statistical analysis leading to parameter restriction. Then, in Section \ref{sec:3}, we discuss the lattice implementation of the model, followed by the results on oscillon formation and the production of gravitational waves. In Section \ref{sec:iv}, we explore the impact of perturbative reheating. Finally, in Section \ref{sec:4}, we present our considerations on the whole analysis and conclude the paper.

\section{The WR model and the new ACT results}\label{sec:2}

The Witten-O'Raifeartaigh (WR) model is given by the potential \cite{ORaifeartaigh:1975nky,Witten:1981kv,Albrecht:1983ib,Martin:2013tda}:
\begin{equation}
    V(\phi) = \Lambda^4\ln^2\left(\frac{\phi}{M}\right),
    \label{eq:2.1}
\end{equation}
motivated by early supersymmetric arguments, but can also be derived from other assumptions within the supergravity framework \cite{Artymowski:2019jlh}. The potential depends on the choice of a mass scale $M$, which determines the minimum of the function, around which the field will oscillate after inflation. In terms of its inflationary observables, the model is seen as disfavored by data, since lower values of $M$, compatible with a lower tensor-to-scalar ratio $r$, lead to a larger scalar spectral index $n_s$, barely within the current CMB constraints. Previous investigations conducted in \cite{Santos:2022aeb,Santos:2023hhk}, where CMB polarization data from BICEP/Keck Array from the 2015 observing season were used, indicate that the inclusion of the more recent 2018 data from the collaboration essentially excludes the model, for a minimal coupling of the field with gravity. 

However, interest in some of these `excluded' models has been regained, since the newest results from the ACT were made available. While being a small-scale survey, significant changes in the constraints for $r$ were not expected; on the other hand, the recent report shows an indication for a flatter primordial power spectrum, in which the analysis in combination with Planck data yielded $n_s=0.9709\pm 0.0038$. The difference becomes even more striking when data from DESI are included in the estimates, reaching $n_s=0.9743\pm 0.0034$, thus motivating the re-analysis of the inflationary predictions for a series of models for the primordial power spectrum. Such a result makes the WR model compelling, since the $n_s>0.97$ region agrees with the predictions for a lower mass scale of the model, while also being interesting from the post-inflationary behavior, as we will see later on.

\begin{figure}
    \centering
    \includegraphics[width=\columnwidth]{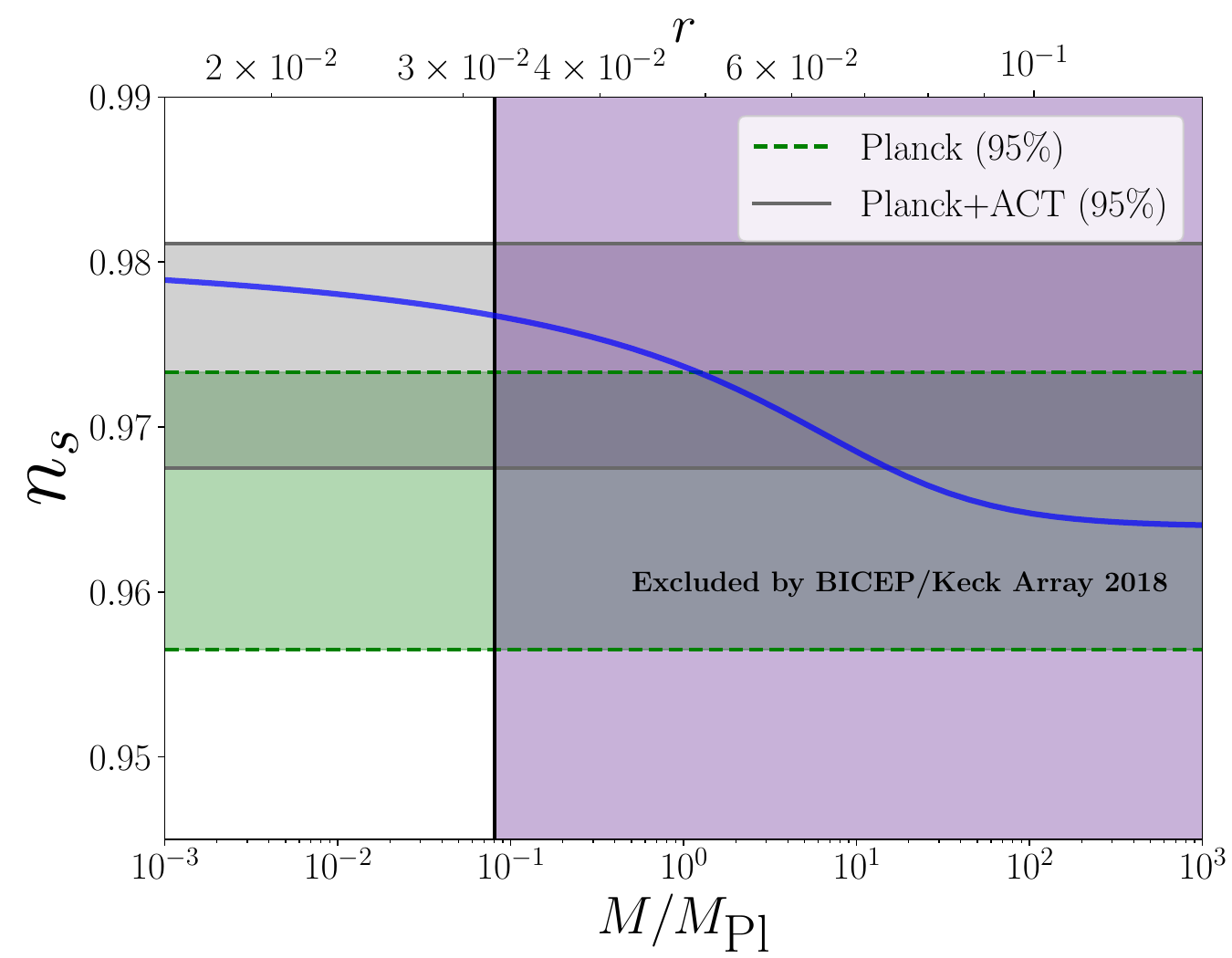}
    \caption{Scalar spectral index $n_s$ as a function of the mass scale $M$ (blue curve). The upper axis shows the corresponding values for the tensor-to-scalar ratio $r$, for which the Planck+BK18 limit is shown as the black vertical line, while the green and gray horizontal bands correspond to the $95\%$ confidence level (C.L.) limits on $n_s$ given by Planck and Planck+ACT, respectively.}
    \label{fig:1}
\end{figure}

Figure \ref{fig:1} shows the model in comparison with current CMB constraints. The blue curve shows the predictions of the WR model for $n_s$ as a function of $M/M_{Pl}$ (in the lower horizontal axis), in which $M_{Pl}$ is the reduced Planck mass. We note that the model is capable of satisfying the limits imposed both by Planck only (green region) and its combination with ACT (gray region), depending on the value of $M$. However, if we consider the constraints on $r$, we notice the following: the upper horizontal axis shows the respective values for $r$, where the black vertical line indicates the current upper limit given by the BICEP/Keck Array (BK18) collaboration, of $r<0.036$. Thus, in order to have the viability of the model, one needs to include the ACT data, for which $M/M_{Pl}\lesssim0.1$ leads to a possible scenario. 

\begin{table}[h!]
\centering
\begin{tabular}{c c} 
 \hline
 Parameter & mean value with 68\% C.L. estimate \\ [0.5ex] 
 \hline\hline
 $\Omega_\mathrm{b} h^2$ & $0.02257\pm 0.00010$ \\
 $\Omega_\mathrm{c} h^2$ & $0.11725\pm 0.00059$ \\
 $\tau_\mathrm{reio}$ & $0.0664\pm 0.0063$  \\
 $\log_{10}M/M_{Pl}$ & $-2.5^{+1.1}_{-1.3}$ \\
 $\log(10^{10} A_\mathrm{s})$ & $3.065\pm 0.012$ \\
 $100\theta$ & $1.04095\pm 0.00026$ \\
 $H_0$ [km s$^{-1}$ Mpc$^{-1}$] & $68.52\pm 0.25$ \\
 $n_s$ & $0.9785^{+0.0013}_{-0.00049}$ \\
 $r$ & $0.0198^{+0.0029}_{-0.0077}$ \\ [1ex] 
\hline
\end{tabular}
\caption{Mean values and uncertainties at 68\% confidence level for the WR model, when considering the analysis of CMB data from Planck with polarization from ACT and BICEP/Keck Array.}
\label{table:1}
\end{table}

The first part of this work will be dedicated to updating the constraints of this model according to Planck+ACT data. Then, we shall use the obtained estimates in our lattice analysis to look into the impact of the model on both the preheating process and the production of post-inflationary relics, such as oscillons. The reheating era can be connected with the inflationary observables, by computing the number of e-folds left until the end of inflation, from the horizon crossing of CMB modes \cite{Planck:2018jri}:
\begin{align}
    N_\ast \simeq 67 - \ln\left(\frac{k_\ast}{a_0H_0}\right) + \frac{1}{4}\ln\left(\frac{V_\ast^2}{M_{Pl}^4V_{\textrm{end}}}\right) + & \nonumber\\ \frac{1-3\bar w}{12(1+\bar w)}\ln\left(\frac{\rho_{\textrm{th}}}{\rho_{\textrm{end}}}\right) - \frac{1}{12}\ln(g_{\textrm{th}}),
    \label{eq:2.2}
\end{align}
with the quantities with `$\ast$' denoting the time of horizon crossing, while the ones with the subscript `$\textrm{th}$' denote the thermalization energy scale, or the beginning of the radiation era. Note that there is a non-negligible dependence on the inflationary model, as well as a weaker dependence on other cosmological parameters, such as the present Hubble expansion rate $H_0$. The ratio $\frac{\rho_{\textrm{th}}}{\rho_{\textrm{end}}}$ encodes the details of reheating, as well as $\bar w$, and might depend on the model for the inflaton decay. We follow \cite{Planck:2018jri}, by noting that for potentials that behave quadratically around the minimum, we can approximate $\bar w\simeq 0$, which is the case for the WR model. Secondly, one can also incorporate a reheating model in the computation of $\frac{\rho_{\textrm{th}}}{\rho_{\textrm{end}}}$, such that the coupling to the inflaton can initially be estimated by data. However, current observations do not lead to sufficiently good restrictions on the inflaton coupling, although the picture might change for future CMB experiments \cite{Drewes:2022nhu}. Therefore, for simplicity, we set $\frac{\rho_{\textrm{th}}}{\rho_{\textrm{end}}}=1$, meaning instant reheating \footnote{This makes the choice of $\bar w$ unimportant in Eq.~ \eqref{eq:2.2}, but we mention this for comparison with our lattice results later on in this work. We fix $k_\star=0.05$ Mpc$^{-1}$ and $g_{\textrm{th}}=10^3$.}. 

\subsection{Methodology and data}

We shall now specify our analysis to be performed in this first part of the work. By quantifying the inflationary behavior of the WR model through its primordial power spectra, we implement this information in the \texttt{CAMB} Boltzmann solver \cite{Lewis:1999bs,Howlett:2012mh}, which, together with the \texttt{cobaya} sampler \cite{2019ascl.soft10019T,Torrado:2020dgo}, allows us to sample the parameter space of the model through the Monte Carlo Markov Chain (MCMC) approach. Along with the usual six $\Lambda$CDM parameters, we vary the mass scale $M$ of the WR model, while the amplitude $\Lambda^4$ is a derived parameter from the restriction given by the amplitude of scalar fluctuations at the CMB scale horizon crossing, $A_s$. As mentioned, the number of e-folds left until the end of inflation $N_\star$ is computed directly from Eq.~\eqref{eq:2.2}, being model-dependent. As for the data used, we will work with the following approach. In order to combine Planck \cite{Planck:2018nkj,Planck:2018vyg} and ACT \cite{Louis_2025} data, one needs to remember that there are overlapping multipoles that require the cutting of the Planck data at some value $\ell$. While we use the low-$\ell$ likelihoods for TT and EE modes, at smaller scales, we follow the ACT team approach to limit $\ell<1000$ for temperature and $\ell<600$ in polarization, achieved by the \texttt{act\_dr6\_cmbonly.PlanckActCut} likelihood, which separates the Planck and ACT contributions. We also add CMB lensing data from the combined surveys, as well as baryonic acoustic oscillations (BAO) observations from the DESI second data release (DESI-DR2) \cite{DESI:2025zgx}. Finally, to obtain the best constraints on $r$, and consequently on $M$, we add data from the CMB B-mode polarization obtained from the BICEP/Keck Array (BK18) \cite{BICEP:2021xfz}.

\subsection{New constraints on the WR model}

\begin{figure*}
    \centering
    \includegraphics[width=1.3\columnwidth]{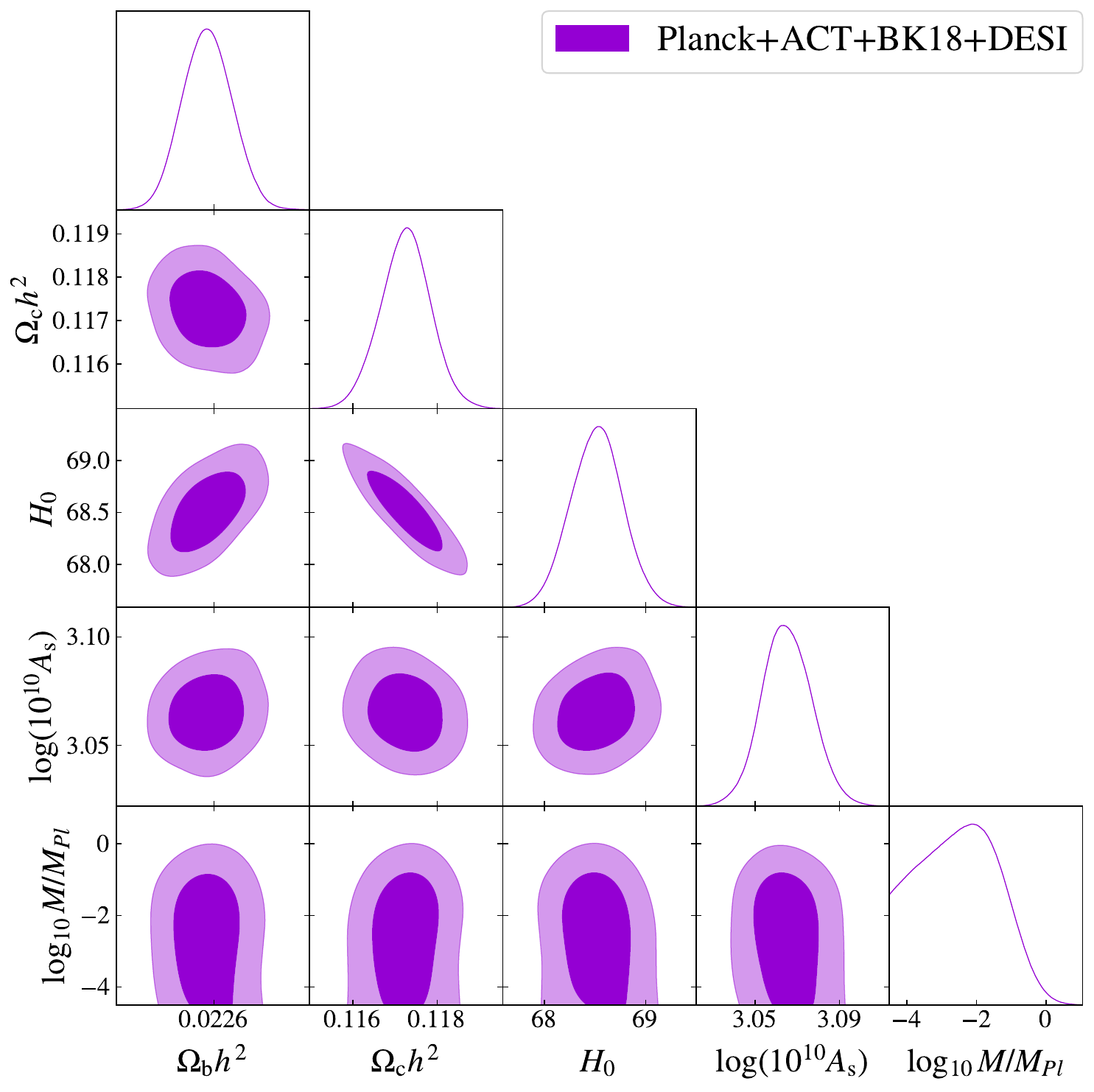}
    \caption{68\% and 95\% confidence level constraints on the WR model, for a Planck+ACT+BK18+DESI analysis.}
    \label{fig:2}
\end{figure*}

\begin{figure}
    \centering
    \includegraphics[width=\columnwidth]{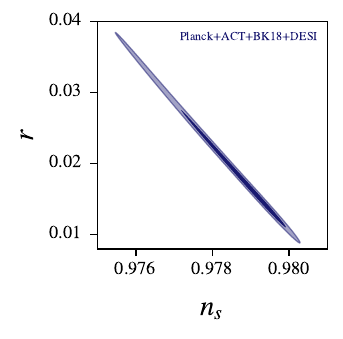}
    \caption{Two-dimensional confidence contours for the $n_s$ and $r$ parameters, as derived for the WR model, at 68\% and 95\% confidence levels.}
    \label{fig:3}
\end{figure}

We expect changes in our results when compared to the work done in \cite{Santos:2022aeb,Santos:2023hhk}, especially when the mass scale $M$ is considered. The results of our statistical analysis are displayed in the Figures \ref{fig:2}, \ref{fig:3} and Table \ref{table:1}. Indeed, we obtain a different constraint on $\log_{10}M$ in which the mass is driven to lower values, such that $\log_{10}M=-2.5^{+1.1}_{-1.3}$. This mean value corresponds to $M/M_{Pl}\simeq 0.00316$, so we get a sub-Planckian value. Comparing this result with the limits in the Figure \ref{fig:1}, we see that the prediction for $n_s$ is well within the gray band that corresponds to the Planck+ACT limits on the $\Lambda$CDM model, while also respecting the upper bound on the tensor-to-scalar ratio $r$. We can see this explicitly by obtaining results for both $n_s$ and $r$ for the model, as derived parameters. We find $n_s=0.9785^{+0.0013}_{-0.00049}$ and $r=0.0198^{+0.0029}_{-0.0077}$; the value for the scalar spectral index is larger than the $\Lambda$CDM estimate achieved by the ACT team, indicating less dependence of the scalar power spectrum with scale within the model. As for the tensor-to-scalar ratio, our estimate is compatible with the current upper limit imposed by the BICEP/Keck Array observatories, at 68\% confidence level. $r$ is also strongly correlated with $n_s$, as seen in Figure \ref{fig:3}; this reflects the behavior of the model, in which large values of $r$ are obtained as $n_s$ decreases. It is important to note that the model is compatible with an already low production of primordial gravitational waves, whose current upper bound should be even more constrained by the next-generation CMB surveys \cite{SimonsObservatory:2018koc,Matsumura:2013aja,LiteBIRD:2022cnt}. If this scenario is confirmed, one can expect even lower values for the mass scale $M$, with a significant impact on the post-inflationary particle production phase in the Universe evolution.

We then see that, in light of the new data, the single-field minimally coupled model is recovered in its viability, in contrast to the one discussed in \cite{Santos:2022aeb,Santos:2023hhk}, in which a non-minimal coupling to gravity had to be considered to achieve this viability. As we will see in the next section, along with concordance with Planck+ACT data, a lower $M$ is interesting from the post-inflationary perspective, since the closer the minimum of the potential is to zero, the more amplified the perturbations during the oscillatory regime will become.

\section{Post-inflationary behavior}\label{sec:3}

After inflation, it is assumed that a period of reheating happens, typically represented as the perturbative decay of the inflaton into relativistic particles of the Standard Model of particle physics. Such a mechanism is achieved if couplings of the inflaton to these particles are allowed after the slow-roll inflationary regime, from which one can predict, based on the intensity of these couplings, the temperature of the thermal bath, or the reheating temperature at which the radiation-dominated universe starts. However, the complete modeling of this period is far from trivial, as limits from Big-Bang nucleosynthesis and possible corrections to the inflationary potential limit the magnitude of the coupling to other particles; also, the first few oscillations of the inflaton around the minimum of the potential can characterize the exponential increase of perturbations in an event called preheating, whose complete characterization can only be achieved through numerically expensive lattice simulations, as discussed in many works \cite{Amin:2010dc,Amin:2011hj,Amin:2014eta,Antusch:2016con,Lozanov:2017hjm,Cotner:2018vug,Antusch:2019qrr,Piani:2023aof,Mahbub:2023faw,Felder:2000hj,Prokopec:1996rr,Copeland:2002ku,Podolsky:2005bw,Garcia-Bellido:2007fiu,Cuissa:2018oiw,Kim:2025ikw,Jia:2024fmo}. In this manner, in the second part of this work, we dismiss the instantaneous reheating approximation to verify the post-inflationary behavior of the WR model, focusing on the self-resonance present in the field dynamics, as well as possible observational consequences, such as the production of oscillons.

\subsection{Equations of motion}

In order to establish the conditions necessary for the lattice study of the model, it is important to focus on the impact of the choice of inflationary parameters and the behavior of the model when the field has its first oscillations around the minimum of the potential. In a FLRW metric, the equation of motion for the field $\phi$ is:
\begin{equation}
    \ddot\phi -a^{-2}\nabla^2\phi + 3H\dot\phi + V_{,\phi} = 0,
    \label{eq:3.1}
\end{equation}
in which $H$ is the Hubble rate of expansion, obtained from the Friedmann equation, as $H^2 = \frac{\rho}{3M_{Pl}^2}$, while $V_{,\phi}$ refers to a derivative of the potential with respect to the scalar field $\phi$. The energy density $\rho_\phi$ and the pressure $P_\phi$ of the field are given as
\begin{equation}
   \rho_\phi = \frac{1}{2}\dot\phi^2 + \frac{1}{2a^2}\nabla^2\phi + V, \quad P_\phi = \frac{1}{2}\dot\phi^2 - \frac{1}{6a^2}\nabla^2\phi - V,
\end{equation}
from which the equation of state $w_\phi=P_\phi/\rho_\phi$ is defined. In the post-inflationary description of the field, we should also look into the perturbations generated as the field is driven towards the minimum. By decomposing the field into a background component $\varphi$ and a fluctuation $\delta\varphi$, direct substitution into Eq.~\eqref{eq:3.1} leads to the known Klein-Gordon background equation, and the evolution of $\delta\varphi$, at first order,
\begin{equation}
    \ddot\varphi + 3H\dot\varphi + V_{,\varphi} = 0, \quad \delta\ddot\varphi -a^{-2}\nabla^2\delta\varphi + 3H\delta\dot\varphi + V_{,\varphi\varphi}\delta\varphi = 0.
    \label{eq:5}
\end{equation}

To solve these equations for a given model, it is customary to rescale some of the variables. In our case, we adopt the following notation,
\begin{equation}
    \chi\equiv\frac{\varphi}{M_{Pl}}, \quad \tau \equiv \frac{\sqrt{\Lambda^4}}{M_{Pl}}t, \quad X \equiv \frac{\sqrt{\Lambda^4}}{M_{Pl}}x, \quad m \equiv \frac{M}{M_{Pl}},
    \label{eq:6}
\end{equation}
as adopted in our $\mathcal{C}$osmo$\mathcal{L}$attice \cite{Figueroa:2020rrl,Figueroa:2021yhd,Figueroa:2023xmq} implementation, with $f_\star=M_{Pl}$ and $\omega_\star=\sqrt{\Lambda^4}/M_{Pl}$. To determine if a model has tachyonic instability, one must examine the second derivative of the potential. As the effective mass of the inflaton field is estimated as (in the redefined units)
\begin{equation}
    m_\chi^2=V_{,\chi\chi}=\frac{2}{\chi^2}(1-\ln(\chi/m)),
\end{equation}
which becomes tachyonic whenever $m_\chi^2(\chi)<0$, yielding $\chi>e^1 \cdot m$ as the region of field space where there can be tachyonic amplification of perturbations. We note that since the potential is not symmetric, tachyonic behavior only happens at the large field regime, and more on this will be discussed later.

\begin{figure}
    \centering
    \includegraphics[width=\columnwidth]{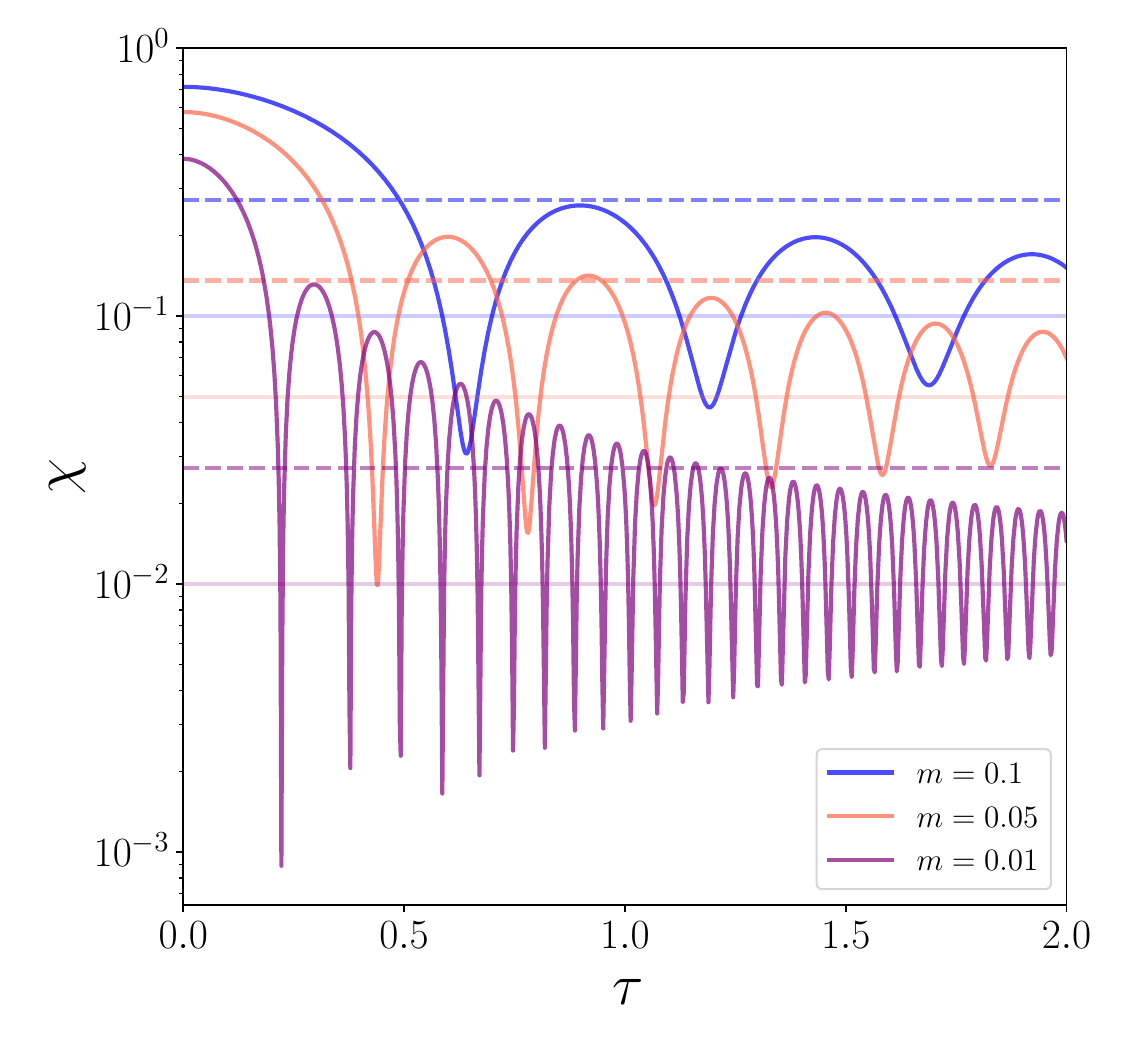}
    \caption{The evolution of the background field $\chi$ for the WR model, as a function of the redefined time $\tau$, for three distinct values of the normalized mass scale $m=0.1$ (blue), $m=0.05$ (red), and $m=0.01$ (purple).}
    \label{fig:4}
\end{figure}

For oscillations around the minimum of the potential, the inflaton mass takes the form,
\begin{equation}
    m_\phi^2 = \frac{2\Lambda^4}{M^2},
\end{equation}
such that the ratio $m_\phi/M_{Pl}$ is around $\sim 2.8\times 10^{-4}$ for $M=0.01 M_{Pl}$, thus being subplanckian. The growth of perturbations of the field is understandable once we work in momentum space. The equation for $\delta\chi$ in Eq.~\eqref{eq:5} becomes
\begin{equation}
    \delta\ddot\chi + 3H\delta\dot\chi + \left(\frac{K^2}{a^2}+V_{,\chi\chi}\right)\delta\chi = 0,
\end{equation}
for a rescaled mode $K$, as given by Eq.~\eqref{eq:6}. A negative mass can lead to negative values in the term between parentheses, for a range of $K$ modes, resulting in an exponentially increasing solution for $\delta\chi$. For the WR model, one can check the impact of choosing different mass scales on the amplification of modes. Figure~\ref{fig:4} shows the background field $\chi$ as a function of $\tau$ for three selected choices of $m$. In the figure, the dashed horizontal lines denote the limit $\chi=e^1 \cdot m$, that might be crossed by the field as it enters or leaves the tachyonic region. We note that as $m$ gets lower, there are more oscillations of the field crossing into the tachyonic region, thus having a larger probability of mode amplification. It is noticeable from the figure that when $m=0.1$, the field never goes back into this region as it starts oscillating; on the other hand, when $m=0.01$, the field oscillates around 11 times before leaving the region definitively. This establishes a correlation between the results obtained in the previous section and the possible production of relics from the tachyonic amplification process. This amplification is quantifiable by solving for the quantity $|\delta\chi_K(\tau=2)/\delta\chi_K(\tau=0)|$, obtained from the solutions of Eq.~\eqref{eq:5}, similarly as done in \cite{Dux:2022kuk} (specifically in their figure 2). By computing this ratio across a large interval of $K$, of $0.1<K<1000$, we find, for instance, that for the $m=0.01$ case, there is a peak in the distribution obtained around $K=100$, helping us to define the momenta to be captured in our simulations; on the other hand, for $m=0.05$, since the field only goes back into the tachyonic region once, we find only a slight amplification of modes at $K\sim 10$, before decreasing and stabilizing. This leads us to understanding the behavior of the $m=0.1$ case, in which there is no amplification at all, and the ratio of perturbations of the field only oscillates near the initial value. For our simulations, we take $N=128^3$ to define the resolution in the lattice, while setting $K_{\textrm{IR}}=6$, being the infrared limit on the wavelength which is related to the size of the lattice box; we have checked that this value is enough to capture all relevant modes for the cases investigated in our study.

\subsection{Simulation results}

\begin{figure}
    \centering
    \includegraphics[width=\columnwidth]{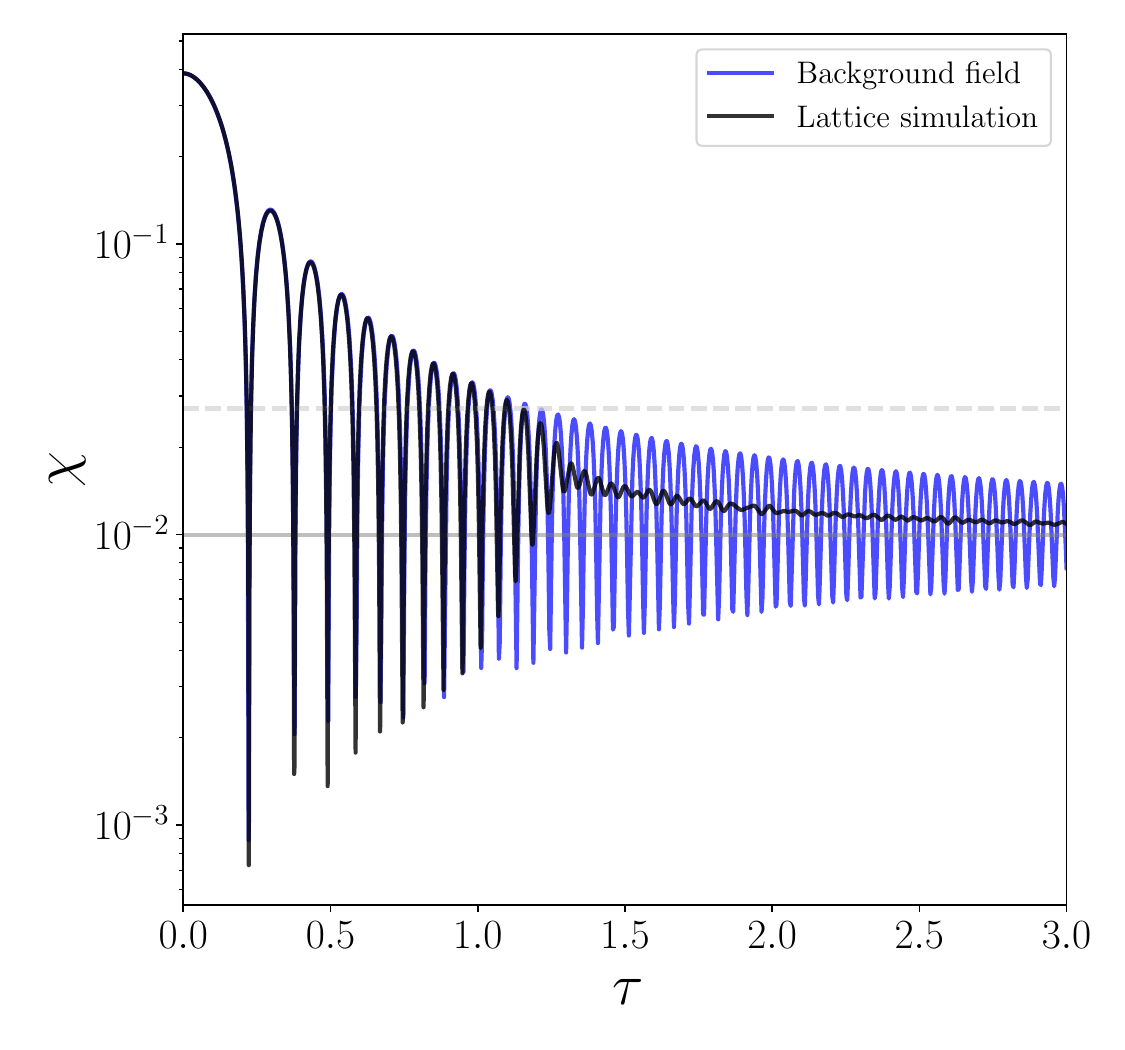}
    \caption{The background field evolution of the WR model (blue), compared with the lattice solution (black), for $m=0.01$. The dashed and solid lines denote the tachyonic region limit and the minimum of the potential, respectively.}
    \label{fig:5}
\end{figure}

\begin{figure}
    \centering
    \includegraphics[width=\columnwidth]{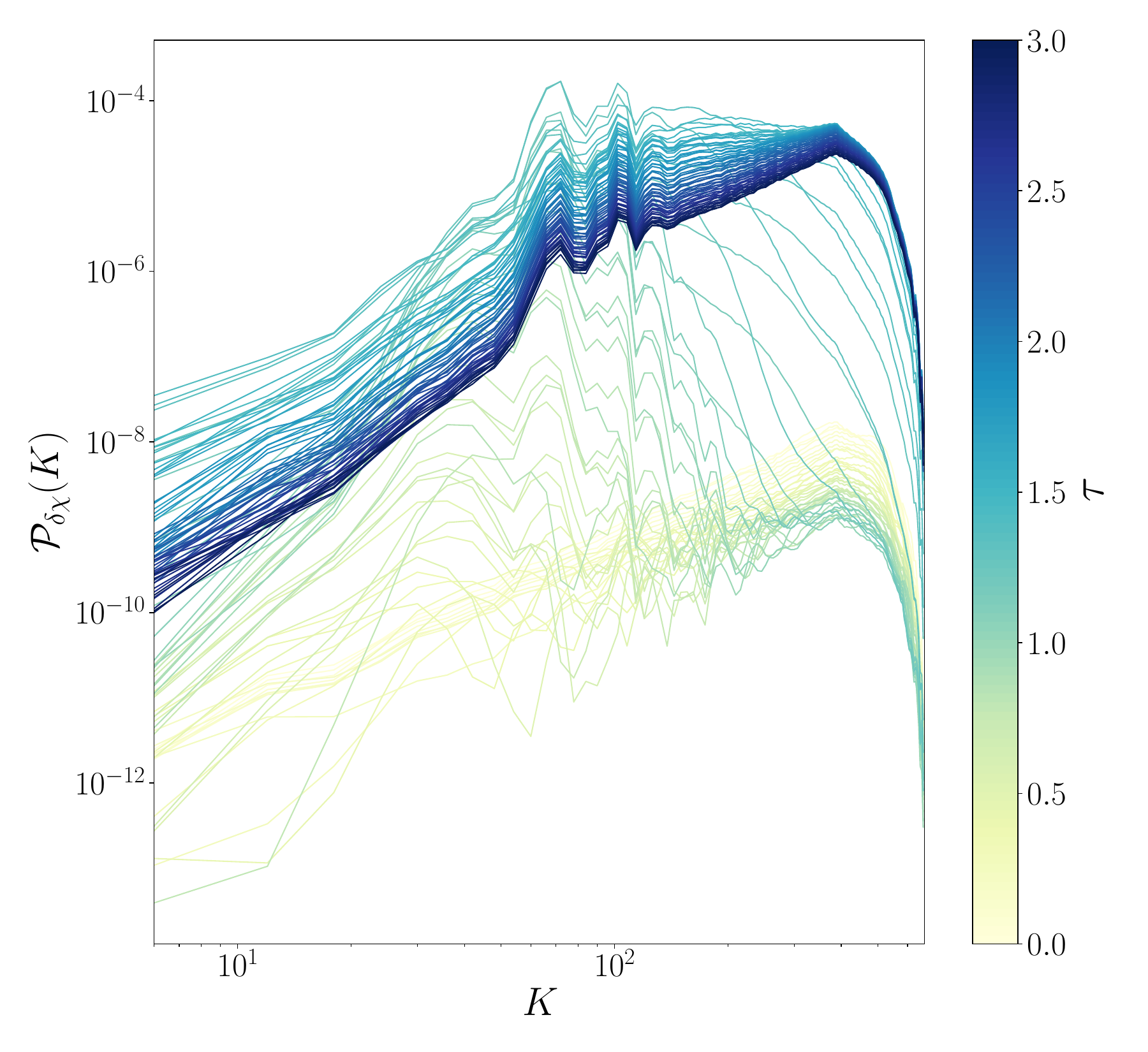}
    \caption{Power spectrum of fluctuations for the WR model with $m=0.01$, as a function of the rescaled wave vector $K$.}
    \label{fig:6}
\end{figure}

We summarize our results in figures \ref{fig:5} and \ref{fig:6}. Figure \ref{fig:5} shows how the field oscillatory behavior differs in the lattice; during the first oscillations around the minimum $m$, the field exits and re-enters the tachyonic region up to around $\tau\sim 1.2$ of the simulation time. During the same time, fluctuations of the field get amplified by tachyonic resonance, and the growth of inhomogeneities is reflected in the increase of the gradient energy $E_G$. The fact that the WR potential is asymmetric, allowing amplification on only one side of $V(\chi)$, plays a significant role in this process, reflected in the number of oscillations necessary for the amplification of $\delta\chi$. After that, the known process of back-reaction into the background field occurs, causing the field oscillations to fragment, seen in the breakdown of the oscillatory behavior.

Figure \ref{fig:6} shows the power spectrum of the fluctuations $\delta\chi$, as a function of the momenta $K$. The colorbar in the figure shows the evolution of the spectra over time. Our initial estimates, derived from solving the model’s equations of motion, indicate that modes in the range 
$5<K<200$ experience the strongest amplification and are amplified earlier. This is clearly observed by comparing the curves in Figure \ref{fig:6} with the corresponding color bars. By the end of the simulation, at $\tau =3$, one sees a shift in the curves in the plot, such that the spectrum has a peak at a larger $K$.

\subsection{Oscillon formation in the model}

\begin{figure*}
    \centering
    \includegraphics[width=\columnwidth]{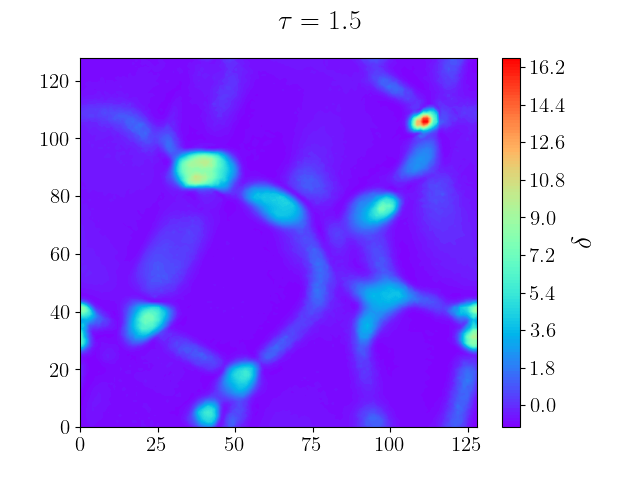}
    \includegraphics[width=\columnwidth]{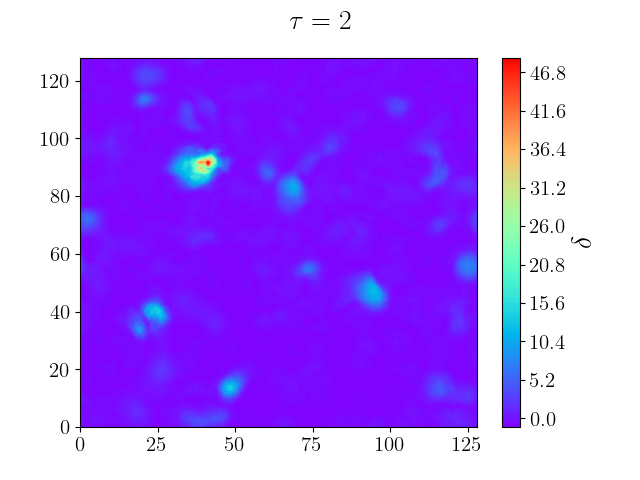}
    \includegraphics[width=\columnwidth]{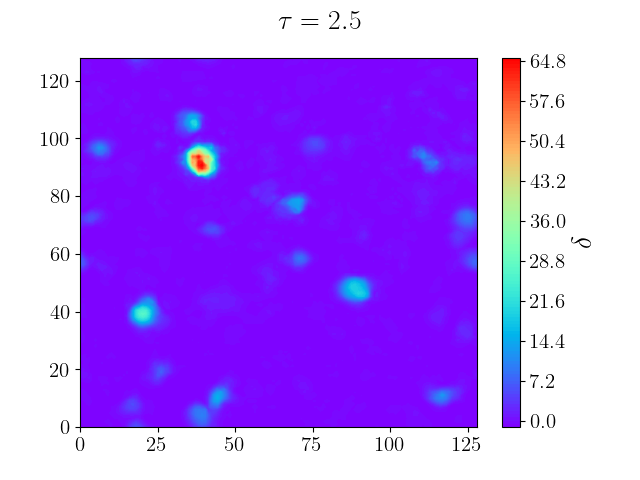}
    \includegraphics[width=\columnwidth]{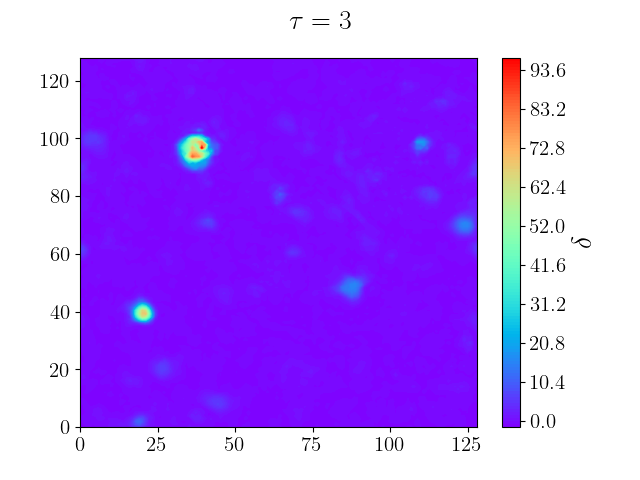}    
    \caption{Snapshots of the lattice simulation for the WR model for $m=0.01$, showing the two-dimensional surfaces of the energy contrast $\delta$, relating the total energy density with the average one. We consider four different times ($\tau=1.5,2,2.5,3$), each of them showing the values for $\delta$ at the color bar at the right of each plot.}
    \label{fig:7}
\end{figure*}

Given the nature of the model, we note that the basic conditions for oscillon formation are met, such that the potential can be approximated as quadratic around the minimum, while it becomes shallower for regions far from it. For the WR model, however, due to the mentioned asymmetry of $V(\phi)$, only one side meets the latter criterion; however, this does not prevent the model from accommodating the creation of oscillons, as seen, for instance, in \cite{Mahbub:2023faw}, for a similar scenario. As $\mathcal{C}$osmo$\mathcal{L}$attice allows saving the 3D distribution of energies throughout the simulation, one can compare different moments in the simulation from the moment the field gets fragmented until the end of the simulation at $\tau=3$. By computing the relative densities $\delta(x)\equiv(\rho(x)-\langle\rho\rangle)/\langle\rho\rangle$, one can localize regions in the lattice for which the density is high with respect to the surroundings. Figure \ref{fig:7} shows two-dimensional snapshots of the 3D energy distribution at four different times, starting around the time the field oscillation gets significantly fragmented, at $\tau=1.5$.

At $\tau=1.5$, we can already see the impact of nonlinearities in the formation of structures, at least for larger scales, before the instabilities get more localized at $\tau=2$. In this case, the red regions correspond to a high density contrast, meaning that almost all of the densities are located in the smaller regions formed. This becomes more evident for $\tau=2.5$ and $\tau=3$, where the number of highly-localized dense regions decreases to a few, but with a high density contrast. This is compatible with oscillon behavior, in which they get smaller and denser with time, as well as acquiring a quasi-spherical distribution.

\subsection{Production of gravitational waves from oscillons}

Perturbations in the FLRW metric allow the production of gravitational waves (GW) $h_{ij}$, which are transverse and traceless, and follow the equation of motion
\begin{equation}
    \ddot h_{ij}^{\textrm{TT}} + 3H\dot h_{ij}^{\textrm{TT}} - \frac{k^2}{a^2} h_{ij}^{\textrm{TT}} = 16\pi G\Pi_{ij}^{\textrm{TT}}, \label{eq:10}
\end{equation}
for the two degrees of freedom achieved from the transverse-traceless (\textrm{TT}) gauge condition. $\Pi_{ij}$ characterizes the source of gravitational waves, the anisotropic tensor, defined as $\Pi_{ij}=T_{ij}-P g_{ij}$. In the usual inflationary picture, the anisotropic tensor is zero at the linear perturbative level, as seen for instance in the case of a scalar field, for which it becomes
\begin{equation}
    \Pi_{ij}^{\textrm{TT}} = \left(\partial_i\phi\partial_j\phi\right)^{\textrm{TT}},
\end{equation}
in the TT gauge. However, if perturbations of the inflaton field become relevant in the non-linear regime, as is the case of a fragmented inflaton under tachyonic resonance, one can link the increase of fluctuations with a potential source of gravitational waves in the early Universe. Since the entering into the non-linear regime and the subsequent fragmentation of the field are related to the formation of oscillons, the production of a sizable gravitational wave background can be expected as a result \cite{Zhou:2013tsa,Liu:2017hua,Antusch:2017vga,Lozanov:2019ylm,Hiramatsu:2020obh,Lozanov:2022yoy,Drees:2025iue,Ballesteros:2024eee}.

\begin{figure*}
    \centering
    \includegraphics[width=1.5\columnwidth]{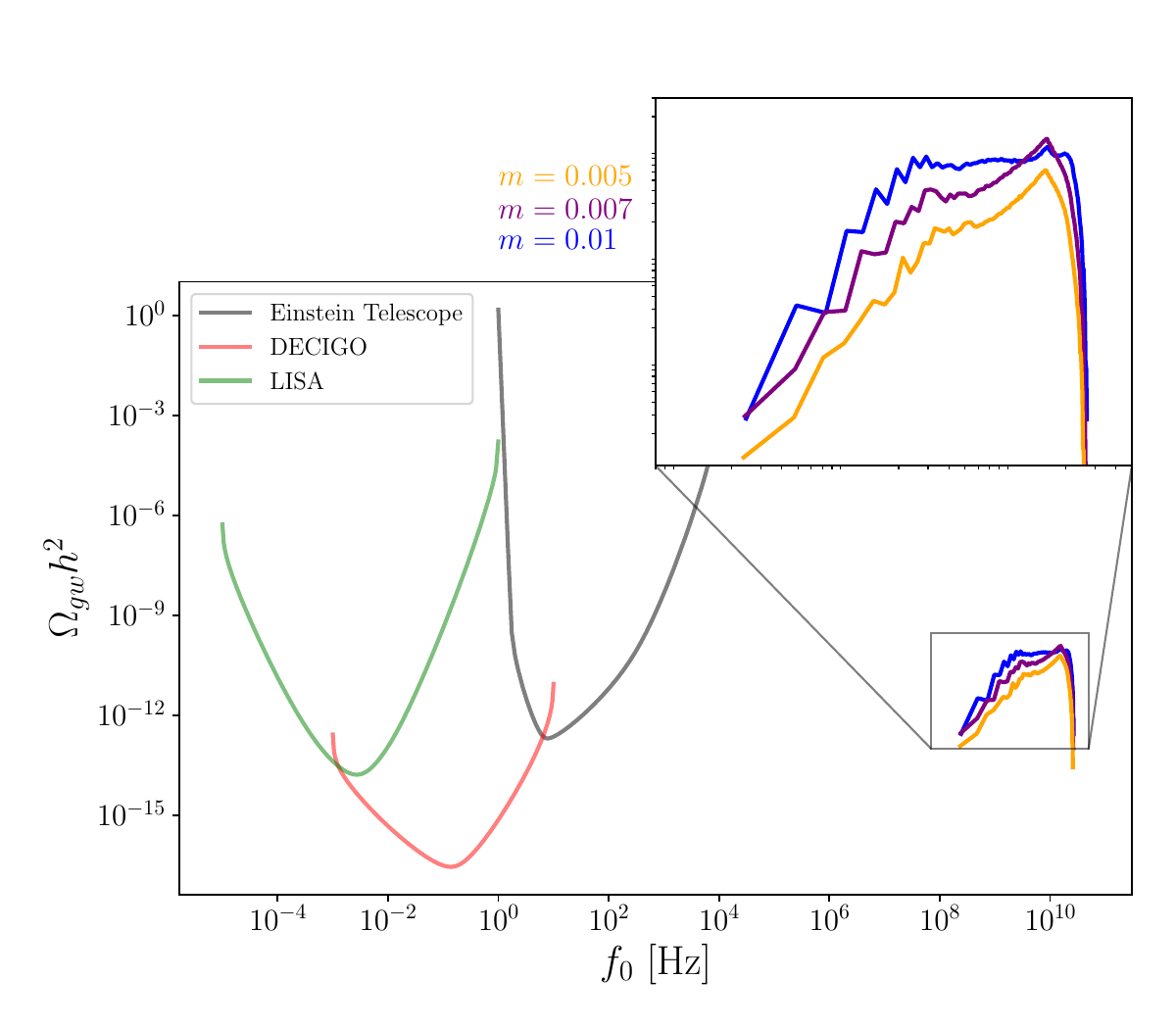}
    \caption{The GW physical density spectrum with frequency $f_0$ for the WR model with $m=0.01$ (blue) compared to the predicted sensitivity of some of the main future detectors for the next generation, being the Einstein Telescope (gray), DECIGO (red), and LISA (green) \cite{Schmitz:2020syl}.}
    \label{fig:8}
\end{figure*}

The energy density of GWs can be computed as solutions to Eq.~\eqref{eq:10} in the second-order perturbative level, as
\begin{equation}
    \rho_{\textrm{GW}} = \frac{M_P^2}{4}\langle\dot h_{ij}\dot h^{ij}\rangle^{\textrm{TT}},
\end{equation}
from which one can compute its contribution to the energy fraction $\Omega_{\textrm{GW}}h^2$ over the cosmic time $\Omega_{\textrm{GW}}h^2(t)=\int \Omega_{\textrm{GW}}h^2(k,t)d\log k$, where
\begin{equation}
    \Omega_{\textrm{GW}}h^2 = \frac{1}{\rho_c}\frac{d\rho_{\textrm{GW}}}{d\log k}.
\end{equation}
As soon as the GWs are produced, they start traveling through space, redshifting as radiation; For a GW associated with a wave vector $k$, the corresponding frequency is given by $f_0=\frac{a}{a_0}\frac{k}{2\pi}$, such that the abundance of gravitational waves today is orders of magnitude smaller than when they were produced \cite{Drees:2025iue}:
\begin{equation}
    \Omega_{\textrm{GW},0}h^2 \sim 10^{-6}\Omega_{\textrm{GW},e}h^2,
\end{equation}
$\Omega_{\textrm{GW},e}$ being the amplitude of GWs when they were emitted. The frequencies $f_0$ observed today can be estimated as \cite{Lozanov:2019ylm,Drees:2025iue}:
\begin{equation}
\begin{aligned}
        f_0 &\simeq  4\times 10^{10}\frac{k}{a_e\rho_e^{1/4}},\\
            &\simeq 4\times 10^{10}\frac{K\Lambda}{a_e\log^{1/4}(\phi_e/M)M_{Pl}},
\end{aligned}
\end{equation}
where we take $\phi_e$ as the beginning of our simulation, with $a_e=1$. Figure \ref{fig:8} shows the GW spectra for $m=0.01,0.007$ and $0.005$ when compared to the future sensitivities of interferometers for the next generation. For the WR model, we obtain results similar to other models \cite{Lozanov:2019ylm,Drees:2025iue}, in which the predicted gravitational wave abundance is well within the future limits; however, the frequencies $f_0$ are too high for any possible detection. Indeed, the peak of the spectrum is located at $f_0\sim 10^9$ Hz, a limit only achievable by some high-frequency detectors, for which possible ideas are a topic of current discussion. Moreover, we notice that the amplitude of the spectra seems to decrease with $M$, for the considered range.

\section{The impact of perturbative reheating}
\label{sec:iv}

In a more general scenario, couplings acquired by the inflation in the post-inflationary epoch can be characterized by the following effective potential:
\begin{equation}
    V(\phi,\chi) = V(\phi) + \frac{1}{2}g^2\phi^2\sigma^2 + \frac{1}{2}y\phi\sigma^2 + \frac{1}{4}\lambda\sigma^2,
\end{equation}
where $V(\phi)$ is the inflaton potential, $g$ denotes the strength of the quadratic-quadratic interaction, while $y$ is the strength of the trilinear interaction. Finally, $\lambda$ is the self-interaction of the potential for $\sigma$, taken as a bosonic field for which the inflaton field can decay to after inflation. During the preheating process, $\phi\phi\rightarrow\sigma\sigma$ is significant through parametric resonance but loses relevance to the $\phi\rightarrow\sigma\sigma$ decay as the perturbative reheating epoch approaches, thus being relevant only in the preheating era. Moreover, a strong quadratic-quadratic coupling can make tachyonic resonance of the fields difficult as discussed in \cite{Antusch:2025ewc}.

In the presence of couplings between the inflaton and other fields, it is possible to estimate the reheating temperature, given the decay width $\Gamma$ of the process. By assuming that the particles produced behave as radiation, such that $\rho_r=\pi^2g_\star T^4_{r}/30$, perturbative reheating becomes relevant when the Hubble expansion rate equals $\Gamma$. Considering the decay of the inflaton to radiation as an instantaneous process,
\begin{equation}
    T_{re} = \sqrt{2}g_\star^{-1/4}\Gamma^{1/2}M_p^{1/2}.
    \label{eq:17}
\end{equation}
Here, $g_\star$ corresponds to the relativistic degrees of freedom of the beginning of the radiation epoch, while $T_{re}$ is denoted as the reheating temperature. In the case of a trilinear decay of the inflaton, we have that
\begin{equation}
    \Gamma = \frac{y^2}{32\pi m_\phi},
    \label{eq:18}
\end{equation}
allowing one to easily compute $T_{re}$ for a period dominated by perturbative reheating. 

We follow the approach of \cite{Ueno:2016dim,Drewes:2017fmn,Drewes:2023bbs}, in which the expected duration of the reheating process $N_{re}$ can be related to the number of e-folds left until the end of inflation $N_k$:
\begin{equation}
    \begin{aligned}
        N_{re} &= \frac{4}{3\bar w_{re}-1}\Bigg[N_k + \ln\left(\frac{k}{a_0 T_0}\right) + \frac{1}{4}\ln\left(\frac{40}{g_\star\pi^2}\right) + \\& \frac{1}{3}\ln\left(\frac{11 g_{s,\star}}{43}\right) - \frac{1}{2}\ln\left(\frac{\pi^2M_{Pl}^2rA_s}{2\sqrt{V_{\textrm end}}}\right)\Bigg],
    \end{aligned}
\end{equation}
with $k=0.05$Mpc$^{-1}$ being the pivot scale of CMB, $a_0=1$ and $T_0=2.725$ K are the current scale factor and the CMB temperature; $g_\star$ and $g_{s,\star}$ are the relativistic degrees of freedom at reheating, which we assume as $g_\star=g_{s,\star}=100$. We note that the model dependence comes in the last term, for which we need to inform the tensor-to-scalar ratio $r$ and the potential $V(\phi)$ at the end of inflation. By assuming that the temperature of the thermal bath generated follows $\rho_{re}=\frac{g_\star\pi^2}{30}T^4_{re}$, one eventually reaches:
\begin{equation}
    T_{re} = \exp\left[-\frac{3(1+\bar w_{re})}{4}N_{re}\right]\left(\frac{40V_{\textrm end}}{g_\star\pi^2}\right)^{1/4}.
\end{equation}

\begin{figure*}
    \centering
    \includegraphics[width=1.3\columnwidth]{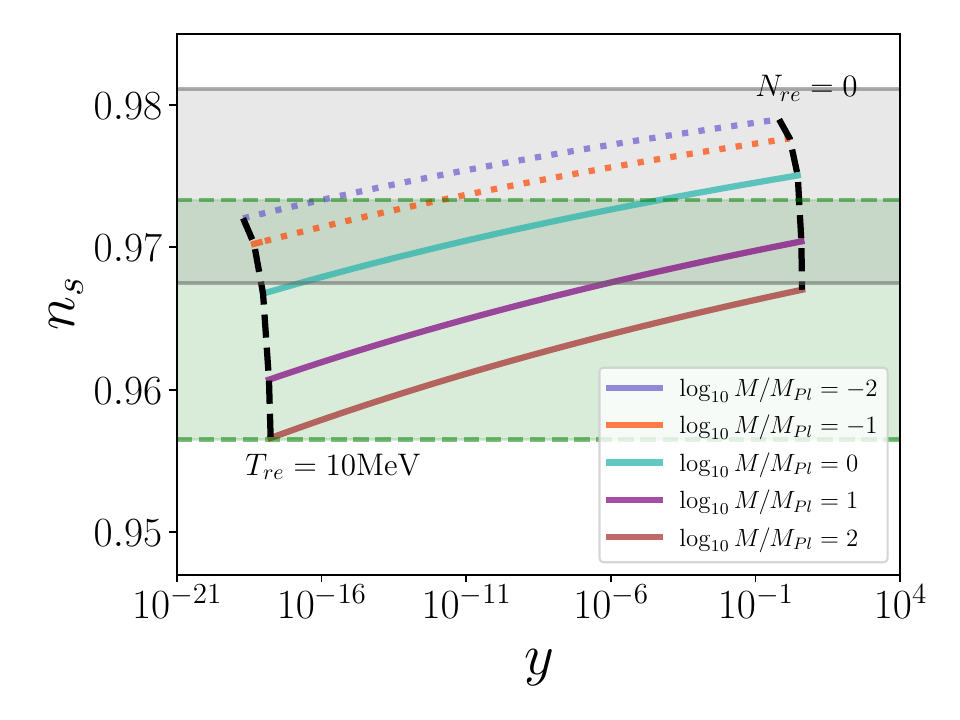}
    \caption{The scalar spectral index $n_s$ as a function of the coupling to bosons $y$, for the WR model with selected values of $M$, as shown by the colored curves, for which the solid and dotted curves represent a corresponding value of the tensor-to-scalar ratio $r$ above and below the BK18 limit, $r<0.038$. The dashed black curves represent the lower limit on the reheating temperature imposed by the BBN and the limit for instantaneous reheating.}
    \label{fig:9}
\end{figure*}

We provide a more general picture of the relation between the coupling strength and the potential parameters as follows. Figure \ref{fig:9} shows the predictions of the spectral index $n_s$ as a function of the coupling strength $y$ in Eq.~\eqref{eq:18}. We select a few values for $\log_{10}M/M_{Pl}$ compatible with either Planck or Planck+ACT to determine the given limits on $y$ according to reheating quantities. Assuming $\bar w_{re}=0$ as the averaged equation of state parameter, we establish the upper values on the coupling from the requirement that the duration of reheating should not give $N_{re}<0$, since $N_{re}=0$ gives the instant reheating case. The coupling strength should then be no more than order one in this case. A lower limit on $y$ can also be obtained by imposing a minimum reheating temperature at which the process can take place, without compromising processes such as the Big Bang Nucleosynthesis (BBN). Adopting the usual value of $T_{\textrm{BBN}}=10$ MeV, we find that the coupling should be at least $y\sim 10^{-18}$ for the WR model. It is also important to relate these results to the severe constraints given by BK18 on the tensor-to-scalar ratio $r$. Values of $n_s$ that obey $r<0.036$ are shown as dotted curves and, as already noticed in our numerical analysis, are consistent with lower values of the mass scale of the model.

Current CMB data do not restrict interactions such as in Eq.~\eqref{eq:18} well, but answers in this direction are expected for next-generation surveys, such as LiteBIRD and Simons Observatory. Forecasts were conducted recently in \cite{Drewes:2022nhu,Drewes:2023bbs}, showing a decrease in several orders of magnitude in the distribution of $y$, for a CMB-S4+LiteBIRD configuration. Thus, the quality of future data sets, combined with the already restrictive theoretical constraints on $N_k$, as seen in Figure~\ref{fig:9}, should finally give us answers on the nature of inflaton couplings during reheating.

\section{Discussion and conclusions}\label{sec:4}

The ACT preference for a higher scalar spectral index, $n_s$, plays a decisive role in revitalizing the inflationary scenarios whose predictions were previously marginal or disfavored. In this work, we revisited the Witten-O'Raifeartaigh inflationary model in light of the current CMB observations, with particular focus on the impact of the ACT DR6 data when combined with Planck and BICEP/Keck Array measurements. This model is interesting for its simplicity, being only dependent on one main parameter, a mass scale $M$ that significantly impacts the viable inflationary parameter space. The fact that the model is essentially discarded given the current Planck data motivates some extensions, such as the inclusion of non-minimal couplings of the inflaton to gravity \cite{Santos:2022aeb,Santos:2023hhk}, and more recently, the application to the warm inflation picture \cite{Das:2025teu}. On the other hand, the consideration of the recent ACT results leads to a compatibility of the model with CMB constraints in its simplest form, a single scalar field minimally coupled to gravity.

In particular, the parameter inference with Planck+ ACT-DR6 + DESI-DR2 + BK18 data results in a tight constraint for the vacuum scale of the model, $\log_{10}M/M_{Pl} = -2.5^{+1.1}_{-1.3}$ at $68\%$ confidence level, which corresponds to a spectral index in the range $n_s=0.9785^{+0.0013}_{-0.00049}$. This sub-Planckian range for $M$ is consistent with lower values for the tensor to scalar ratio, aligning the slow-roll predictions for $r$ with the upper limit set by the BICEP/Keck Array experiments, $r<0.036$, without resorting to a non-minimal coupling to gravity.

The low-$M$ regime also has a significant impact on the post-inflationary dynamics of the inflaton field, once the structure of the WR potential supports a tachyonic instability during the early oscillatory phase. More specifically, the perturbations in the inflaton field are amplified as the amplitude of the damped oscillations crosses the tachyonic region, $\phi>e^1 \cdot M$. Using lattice simulations, we investigated the self-resonant behavior of the inflaton, explicitly demonstrating the fragmentation of the field condensate and the formation of localized, long-lived configurations consistent with oscillons. After the fragmentation of the field, power is transferred to the smaller scales, leading to the formation of oscillons. Such a process establishes a nontrivial link between CMB observables and the nonlinear dynamics in the early Universe.

The process of fragmentation of the inflaton field and subsequent oscillon formation can also contribute significantly as a source of anisotropic stress and, consequently, to gravitational waves. We computed the predicted spectrum of the gravitational waves generated in the WR model and found that, although the abundance is within the limits of future experiments such as LISA, DECIGO, and Einstein Telescope, the predicted frequencies around $f_0 \sim 10^9\,\,\text{Hz}$ are too high for any possible detection in the near future. This result is consistent with previous studies of oscillon-induced gravitational wave production and highlights the need for dedicated high-frequency detection strategies to probe such signals.

Finally, we investigated the implications of perturbative reheating through a trilinear inflaton-scalar coupling. In particular, we obtained $10^{-18} \lesssim y \lesssim \mathcal{O}(1)$ when considering the limits from the BBN temperature and instantaneous reheating. Overall, our results demonstrate not only the capability of the WR model to describe the recent CMB observations but also a framework with rich and predictable post-inflationary dynamics. This highlights the potential of combining precision cosmology and computational simulations to probe the microphysics of the early Universe.

\section*{Acknowledgements}

We acknowledge the use of high-performance computing services provided by the Observatório Nacional Data Center. FBMS is supported by Conselho Nacional de Desenvolvimento Científico e Tecnológico (CNPq) grant No. 151554/2024-2. JGR is supported by FAPERJ grant No. E-26/200.513/2025 and CNPq grant No. 406718/2025-3. GR is supported by the Coordena\c{c}\~ao de Aperfei\c{c}oamento de Pessoal de N\'ivel Superior (CAPES). CS is supported by CNPq through grant numbers 304944/2025-4 and 406718/2025-3 and by the São Paulo Research Foundation (FAPESP) through grant number 2021/01089-1. JSA is supported by CNPq No. 307683/2022-2 and FAPERJ grant 299312 (2023).

\bibliography{references}

@article{Starobinsky:1980te,
    author = "Starobinsky, Alexei A.",
    editor = "Khalatnikov, I. M. and Mineev, V. P.",
    title = "{A New Type of Isotropic Cosmological Models Without Singularity}",
    doi = "10.1016/0370-2693(80)90670-X",
    journal = "Phys. Lett. B",
    volume = "91",
    pages = "99--102",
    year = "1980"
}

@article{Guth:1980zm,
    author = "Guth, Alan H.",
    editor = "Fang, Li-Zhi and Ruffini, R.",
    title = "{The Inflationary Universe: A Possible Solution to the Horizon and Flatness Problems}",
    reportNumber = "SLAC-PUB-2576",
    doi = "10.1103/PhysRevD.23.347",
    journal = "Phys. Rev. D",
    volume = "23",
    pages = "347--356",
    year = "1981"
}

@article{Linde:1981mu,
    author = "Linde, Andrei D.",
    editor = "Fang, Li-Zhi and Ruffini, R.",
    title = "{A New Inflationary Universe Scenario: A Possible Solution of the Horizon, Flatness, Homogeneity, Isotropy and Primordial Monopole Problems}",
    reportNumber = "LEBEDEV-81-229",
    doi = "10.1016/0370-2693(82)91219-9",
    journal = "Phys. Lett. B",
    volume = "108",
    pages = "389--393",
    year = "1982"
}

@article{Albrecht:1982mp,
    author = "Albrecht, Andreas and Steinhardt, Paul J. and Turner, Michael S. and Wilczek, Frank",
    title = "{Reheating an Inflationary Universe}",
    reportNumber = "UPR-0189T, EFI-82-09-CHICAGO",
    doi = "10.1103/PhysRevLett.48.1437",
    journal = "Phys. Rev. Lett.",
    volume = "48",
    pages = "1437",
    year = "1982"
}

@article{Abbott:1982hn,
    author = "Abbott, L. F. and Farhi, Edward and Wise, Mark B.",
    title = "{Particle Production in the New Inflationary Cosmology}",
    reportNumber = "MIT-CTP-983",
    doi = "10.1016/0370-2693(82)90867-X",
    journal = "Phys. Lett. B",
    volume = "117",
    pages = "29",
    year = "1982"
}

@article{Kofman:1994rk,
    author = "Kofman, Lev and Linde, Andrei D. and Starobinsky, Alexei A.",
    title = "{Reheating after inflation}",
    eprint = "hep-th/9405187",
    archivePrefix = "arXiv",
    reportNumber = "UH-IFA-94-35, SU-ITP-94-13, YITP-U-94-15",
    doi = "10.1103/PhysRevLett.73.3195",
    journal = "Phys. Rev. Lett.",
    volume = "73",
    pages = "3195--3198",
    year = "1994"
}

@article{Kofman:1997yn,
    author = "Kofman, Lev and Linde, Andrei D. and Starobinsky, Alexei A.",
    title = "{Towards the theory of reheating after inflation}",
    eprint = "hep-ph/9704452",
    archivePrefix = "arXiv",
    reportNumber = "IFA-97-28, SU-ITP-97-18",
    doi = "10.1103/PhysRevD.56.3258",
    journal = "Phys. Rev. D",
    volume = "56",
    pages = "3258--3295",
    year = "1997"
}

@article{Gleiser:1993pt,
    author = "Gleiser, Marcelo",
    title = "{Pseudostable bubbles}",
    eprint = "hep-ph/9308279",
    archivePrefix = "arXiv",
    reportNumber = "DART-HEP-93-05",
    doi = "10.1103/PhysRevD.49.2978",
    journal = "Phys. Rev. D",
    volume = "49",
    pages = "2978--2981",
    year = "1994"
}

@article{Copeland:1995fq,
    author = "Copeland, Edmund J. and Gleiser, M. and Muller, H. -R.",
    title = "{Oscillons: Resonant configurations during bubble collapse}",
    eprint = "hep-ph/9503217",
    archivePrefix = "arXiv",
    reportNumber = "SUSX-TH-95-3-3, FERMILAB-PUB-95-021-A, DART-HEP-95-01",
    doi = "10.1103/PhysRevD.52.1920",
    journal = "Phys. Rev. D",
    volume = "52",
    pages = "1920--1933",
    year = "1995"
}

@article{Amin:2010jq,
    author = "Amin, Mustafa A. and Shirokoff, David",
    title = "{Flat-top oscillons in an expanding universe}",
    eprint = "1002.3380",
    archivePrefix = "arXiv",
    primaryClass = "astro-ph.CO",
    doi = "10.1103/PhysRevD.81.085045",
    journal = "Phys. Rev. D",
    volume = "81",
    pages = "085045",
    year = "2010"
}

@article{Amin:2010dc,
    author = "Amin, Mustafa A. and Easther, Richard and Finkel, Hal",
    title = "{Inflaton Fragmentation and Oscillon Formation in Three Dimensions}",
    eprint = "1009.2505",
    archivePrefix = "arXiv",
    primaryClass = "astro-ph.CO",
    doi = "10.1088/1475-7516/2010/12/001",
    journal = "JCAP",
    volume = "12",
    pages = "001",
    year = "2010"
}

@article{Amin:2011hj,
    author = "Amin, Mustafa A. and Easther, Richard and Finkel, Hal and Flauger, Raphael and Hertzberg, Mark P.",
    title = "{Oscillons After Inflation}",
    eprint = "1106.3335",
    archivePrefix = "arXiv",
    primaryClass = "astro-ph.CO",
    doi = "10.1103/PhysRevLett.108.241302",
    journal = "Phys. Rev. Lett.",
    volume = "108",
    pages = "241302",
    year = "2012"
}

@article{Amin:2014eta,
    author = "Amin, Mustafa A. and Hertzberg, Mark P. and Kaiser, David I. and Karouby, Johanna",
    title = "{Nonperturbative Dynamics Of Reheating After Inflation: A Review}",
    eprint = "1410.3808",
    archivePrefix = "arXiv",
    primaryClass = "hep-ph",
    doi = "10.1142/S0218271815300037",
    journal = "Int. J. Mod. Phys. D",
    volume = "24",
    pages = "1530003",
    year = "2014"
}

@article{Antusch:2016con,
    author = "Antusch, Stefan and Cefala, Francesco and Orani, Stefano",
    title = "{Gravitational waves from oscillons after inflation}",
    eprint = "1607.01314",
    archivePrefix = "arXiv",
    primaryClass = "astro-ph.CO",
    doi = "10.1103/PhysRevLett.118.011303",
    journal = "Phys. Rev. Lett.",
    volume = "118",
    number = "1",
    pages = "011303",
    year = "2017",
    note = "[Erratum: Phys.Rev.Lett. 120, 219901 (2018)]"
}

@article{Lozanov:2017hjm,
    author = "Lozanov, Kaloian D. and Amin, Mustafa A.",
    title = "{Self-resonance after inflation: oscillons, transients and radiation domination}",
    eprint = "1710.06851",
    archivePrefix = "arXiv",
    primaryClass = "astro-ph.CO",
    doi = "10.1103/PhysRevD.97.023533",
    journal = "Phys. Rev. D",
    volume = "97",
    number = "2",
    pages = "023533",
    year = "2018"
}

@article{Cotner:2018vug,
    author = "Cotner, Eric and Kusenko, Alexander and Takhistov, Volodymyr",
    title = "{Primordial Black Holes from Inflaton Fragmentation into Oscillons}",
    eprint = "1801.03321",
    archivePrefix = "arXiv",
    primaryClass = "astro-ph.CO",
    reportNumber = "IPMU18-0008",
    doi = "10.1103/PhysRevD.98.083513",
    journal = "Phys. Rev. D",
    volume = "98",
    number = "8",
    pages = "083513",
    year = "2018"
}

@article{Antusch:2019qrr,
    author = "Antusch, Stefan and Cefal{\`a}, Francesco and Torrent{\'\i}, Francisco",
    title = "{Properties of Oscillons in Hilltop Potentials: energies, shapes, and lifetimes}",
    eprint = "1907.00611",
    archivePrefix = "arXiv",
    primaryClass = "hep-ph",
    doi = "10.1088/1475-7516/2019/10/002",
    journal = "JCAP",
    volume = "10",
    pages = "002",
    year = "2019"
}

@article{Piani:2023aof,
    author = "Piani, Matteo and Rubio, Javier",
    title = "{Preheating in Einstein-Cartan Higgs Inflation: oscillon formation}",
    eprint = "2304.13056",
    archivePrefix = "arXiv",
    primaryClass = "hep-ph",
    reportNumber = "IPARCOS-UCM-23-031",
    doi = "10.1088/1475-7516/2023/12/002",
    journal = "JCAP",
    volume = "12",
    pages = "002",
    year = "2023"
}

@article{Mahbub:2023faw,
    author = "Mahbub, Rafid and Mishra, Swagat S.",
    title = "{Oscillon formation from preheating in asymmetric inflationary potentials}",
    eprint = "2303.07503",
    archivePrefix = "arXiv",
    primaryClass = "astro-ph.CO",
    doi = "10.1103/PhysRevD.108.063524",
    journal = "Phys. Rev. D",
    volume = "108",
    number = "6",
    pages = "063524",
    year = "2023"
}

@article{Kasai:2025coe,
    author = "Kasai, Kentaro and Kitajima, Naoya",
    title = "{Primordial black hole formation from the merger of oscillons}",
    eprint = "2510.25300",
    archivePrefix = "arXiv",
    primaryClass = "astro-ph.CO",
    doi = "10.1103/9965-5h5h",
    journal = "Phys. Rev. D",
    volume = "113",
    number = "6",
    pages = "L061307",
    year = "2026"
}

@article{Zhou:2013tsa,
    author = "Zhou, Shuang-Yong and Copeland, Edmund J. and Easther, Richard and Finkel, Hal and Mou, Zong-Gang and Saffin, Paul M.",
    title = "{Gravitational Waves from Oscillon Preheating}",
    eprint = "1304.6094",
    archivePrefix = "arXiv",
    primaryClass = "astro-ph.CO",
    doi = "10.1007/JHEP10(2013)026",
    journal = "JHEP",
    volume = "10",
    pages = "026",
    year = "2013"
}

@article{Liu:2017hua,
    author = "Liu, Jing and Guo, Zong-Kuan and Cai, Rong-Gen and Shiu, Gary",
    title = "{Gravitational Waves from Oscillons with Cuspy Potentials}",
    eprint = "1707.09841",
    archivePrefix = "arXiv",
    primaryClass = "astro-ph.CO",
    doi = "10.1103/PhysRevLett.120.031301",
    journal = "Phys. Rev. Lett.",
    volume = "120",
    number = "3",
    pages = "031301",
    year = "2018"
}

@article{Antusch:2017vga,
    author = "Antusch, Stefan and Cefala, Francesco and Orani, Stefano",
    title = "{What can we learn from the stochastic gravitational wave background produced by oscillons?}",
    eprint = "1712.03231",
    archivePrefix = "arXiv",
    primaryClass = "astro-ph.CO",
    doi = "10.1088/1475-7516/2018/03/032",
    journal = "JCAP",
    volume = "03",
    pages = "032",
    year = "2018"
}

@article{Lozanov:2019ylm,
    author = "Lozanov, Kaloian D. and Amin, Mustafa A.",
    title = "{Gravitational perturbations from oscillons and transients after inflation}",
    eprint = "1902.06736",
    archivePrefix = "arXiv",
    primaryClass = "astro-ph.CO",
    doi = "10.1103/PhysRevD.99.123504",
    journal = "Phys. Rev. D",
    volume = "99",
    number = "12",
    pages = "123504",
    year = "2019"
}

@article{Hiramatsu:2020obh,
    author = "Hiramatsu, Takashi and Sfakianakis, Evangelos I. and Yamaguchi, Masahide",
    title = "{Gravitational wave spectra from oscillon formation after inflation}",
    eprint = "2011.12201",
    archivePrefix = "arXiv",
    primaryClass = "hep-ph",
    reportNumber = "Nikhef 2020-028, RUP-20-33",
    doi = "10.1007/JHEP03(2021)021",
    journal = "JHEP",
    volume = "03",
    pages = "021",
    year = "2021"
}

@article{Lozanov:2022yoy,
    author = "Lozanov, Kaloian D. and Takhistov, Volodymyr",
    title = "{Enhanced Gravitational Waves from Inflaton Oscillons}",
    eprint = "2204.07152",
    archivePrefix = "arXiv",
    primaryClass = "astro-ph.CO",
    reportNumber = "IPMU22-0016, KEK-QUP-2023-0008, KEK-TH-2518, KEK-Cosmo-0311",
    doi = "10.1103/PhysRevLett.130.181002",
    journal = "Phys. Rev. Lett.",
    volume = "130",
    number = "18",
    pages = "181002",
    year = "2023"
}

@article{Drees:2025iue,
    author = "Drees, Manuel and Wang, Chenhuan",
    title = "{Inflaton self resonance, oscillons, and gravitational waves in small field polynomial inflation}",
    eprint = "2501.13811",
    archivePrefix = "arXiv",
    primaryClass = "astro-ph.CO",
    doi = "10.1088/1475-7516/2025/04/078",
    journal = "JCAP",
    volume = "04",
    pages = "078",
    year = "2025"
}

@article{Louis_2025,
doi = {10.1088/1475-7516/2025/11/062},
url = {https://doi.org/10.1088/1475-7516/2025/11/062},
year = {2025},
month = {nov},
publisher = {IOP Publishing},
volume = {2025},
number = {11},
pages = {062},
author = {Louis, Thibaut and others (The Atacama Cosmology Telescope collaboration)},
title = {The Atacama Cosmology Telescope: DR6 power spectra, likelihoods and ΛCDM parameters},
journal = {Journal of Cosmology and Astroparticle Physics}
}

@article{Ballesteros:2024eee,
    author = "Ballesteros, Guillermo and Iguaz Juan, Joaquim and Serpico, Pasquale D. and Taoso, Marco",
    title = "{Primordial black hole formation from self-resonant preheating?}",
    eprint = "2406.09122",
    archivePrefix = "arXiv",
    primaryClass = "astro-ph.CO",
    doi = "10.1103/PhysRevD.111.083521",
    journal = "Phys. Rev. D",
    volume = "111",
    number = "8",
    pages = "083521",
    year = "2025"
}

@article{Planck:2018jri,
    author = "Akrami, Y. and others",
    collaboration = "Planck",
    title = "{Planck 2018 results. X. Constraints on inflation}",
    eprint = "1807.06211",
    archivePrefix = "arXiv",
    primaryClass = "astro-ph.CO",
    doi = "10.1051/0004-6361/201833887",
    journal = "Astron. Astrophys.",
    volume = "641",
    pages = "A10",
    year = "2020"
}

@article{DESI:2024mwx,
    author = "Adame, A. G. and others",
    collaboration = "DESI",
    title = "{DESI 2024 VI: cosmological constraints from the measurements of baryon acoustic oscillations}",
    eprint = "2404.03002",
    archivePrefix = "arXiv",
    primaryClass = "astro-ph.CO",
    reportNumber = "FERMILAB-PUB-24-0154-PPD",
    doi = "10.1088/1475-7516/2025/02/021",
    journal = "JCAP",
    volume = "02",
    pages = "021",
    year = "2025"
}

@article{DESI:2025zgx,
    author = "Abdul Karim, M. and others",
    collaboration = "DESI",
    title = "{DESI DR2 results. II. Measurements of baryon acoustic oscillations and cosmological constraints}",
    eprint = "2503.14738",
    archivePrefix = "arXiv",
    primaryClass = "astro-ph.CO",
    reportNumber = "FERMILAB-PUB-25-0169-PPD",
    doi = "10.1103/tr6y-kpc6",
    journal = "Phys. Rev. D",
    volume = "112",
    number = "8",
    pages = "083515",
    year = "2025"
}

@article{Zharov:2025zjg,
    author = "Zharov, D. S. and Sobol, O. O. and Vilchinskii, S. I.",
    title = "{ACT observations, reheating, and Starobinsky and Higgs inflation}",
    doi = "10.1103/km3q-rm34",
    journal = "Phys. Rev. D",
    volume = "112",
    number = "2",
    pages = "023544",
    year = "2025"
}

@article{lq71:b84v,
  title = {BAO-CMB tension and implications for inflation},
  author = {Ferreira, Elisa G. M. and McDonough, Evan and Balkenhol, Lennart and Kallosh, Renata and Knox, Lloyd and Linde, Andrei},
  journal = {Phys. Rev. D},
  pages = {--},
  year = {2025},
  month = {Dec},
  publisher = {American Physical Society},
  doi = {10.1103/lq71-b84v},
  url = {https://link.aps.org/doi/10.1103/lq71-b84v}
}

@article{Haque:2025uri,
    author = "Haque, Md Riajul and Pal, Sourav and Paul, Debarun",
    title = "{ACT DR6 Insights on the Inflationary Attractor models and Reheating}",
    eprint = "2505.01517",
    journal = {},
    archivePrefix = "arXiv",
    primaryClass = "astro-ph.CO",
    month = "5",
    year = "2025"
}

@article{Peng:2025bws,
    author = "Peng, Zhi-Zhang and Chen, Zu-Cheng and Liu, Lang",
    title = "{Polynomial potential inflation in the ACT era: From CMB to primordial black holes}",
    eprint = "2505.12816",
    archivePrefix = "arXiv",
    primaryClass = "astro-ph.CO",
    doi = "10.1103/hzcf-q2rk",
    journal = "Phys. Rev. D",
    volume = "113",
    number = "6",
    pages = "063527",
    year = "2026"
}

@article{Addazi:2025qra,
    author = "Addazi, Andrea and Aldabergenov, Yermek and Ketov, Sergei V.",
    title = "{Curvature corrections to Starobinsky inflation can explain the ACT results}",
    eprint = "2505.10305",
    archivePrefix = "arXiv",
    primaryClass = "gr-qc",
    reportNumber = "IPMU25-0021",
    doi = "10.1016/j.physletb.2025.139883",
    journal = "Phys. Lett. B",
    volume = "869",
    pages = "139883",
    year = "2025"
}

@article{Berera:2025vsu,
    author = "Berera, Arjun and Brahma, Suddhasattwa and Qiu, Zizang and O. Ramos, Rudnei and Rodrigues, Gabriel S.",
    title = "{The early universe is ACT-ing warm}",
    eprint = "2504.02655",
    archivePrefix = "arXiv",
    primaryClass = "hep-th",
    doi = "10.1088/1475-7516/2025/11/059",
    journal = "JCAP",
    volume = "11",
    pages = "059",
    year = "2025"
}

@article{Gao:2025onc,
    author = "Gao, Qing and Gong, Yungui and Yi, Zhu and Zhang, Fengge",
    title = "{Nonminimal coupling in light of ACT data}",
    eprint = "2504.15218",
    archivePrefix = "arXiv",
    primaryClass = "astro-ph.CO",
    doi = "10.1016/j.dark.2025.102106",
    journal = "Phys. Dark Univ.",
    volume = "50",
    pages = "102106",
    year = "2025"
}

@article{Yogesh:2025wak,
    author = "Yogesh and Mohammadi, Abolhassan and Wu, Qiang and Zhu, Tao",
    title = "{Starobinsky like inflation and EGB Gravity in the light of ACT}",
    eprint = "2505.05363",
    archivePrefix = "arXiv",
    primaryClass = "astro-ph.CO",
    doi = "10.1088/1475-7516/2025/10/010",
    journal = "JCAP",
    volume = "10",
    pages = "010",
    year = "2025"
}

@article{Liu:2025qca,
    author = "Liu, Lang and Yi, Zhu and Gong, Yungui",
    title = "{Reconciling Higgs Inflation with ACT Observations through Reheating}",
    eprint = "2505.02407",
    journal = {},
    archivePrefix = "arXiv",
    primaryClass = "astro-ph.CO",
    month = "5",
    year = "2025"
}

@article{Drees:2025ngb,
    author = "Drees, Manuel and Xu, Yong",
    title = "{Refined predictions for Starobinsky inflation and post-inflationary constraints in light of ACT}",
    eprint = "2504.20757",
    archivePrefix = "arXiv",
    primaryClass = "astro-ph.CO",
    reportNumber = "MITP-25-033",
    doi = "10.1016/j.physletb.2025.139612",
    journal = "Phys. Lett. B",
    volume = "867",
    pages = "139612",
    year = "2025"
}

@article{Kallosh:2025rni,
    author = "Kallosh, Renata and Linde, Andrei and Roest, Diederik",
    title = "{Atacama Cosmology Telescope, South Pole Telescope, and Chaotic Inflation}",
    eprint = "2503.21030",
    archivePrefix = "arXiv",
    primaryClass = "hep-th",
    doi = "10.1103/d6gn-78hn",
    journal = "Phys. Rev. Lett.",
    volume = "135",
    number = "16",
    pages = "161001",
    year = "2025"
}

@article{ORaifeartaigh:1975nky,
    author = "O'Raifeartaigh, L.",
    title = "{Spontaneous Symmetry Breaking for Chiral Scalar Superfields}",
    reportNumber = "DIAS-TP-75-9",
    doi = "10.1016/0550-3213(75)90585-4",
    journal = "Nucl. Phys. B",
    volume = "96",
    pages = "331--352",
    year = "1975"
}

@article{Witten:1981kv,
    author = "Witten, Edward",
    title = "{Mass Hierarchies in Supersymmetric Theories}",
    reportNumber = "IC/81/106",
    doi = "10.1016/0370-2693(81)90885-6",
    journal = "Phys. Lett. B",
    volume = "105",
    pages = "267",
    year = "1981"
}

@article{Albrecht:1983ib,
    author = "Albrecht, Andreas and Dimopoulos, Savas and Fischler, W. and Kolb, Edward W. and Raby, Stuart and Steinhardt, Paul J.",
    title = "{New Inflation in Supersymmetric Theories}",
    reportNumber = "LA-UR-83-597",
    doi = "10.1016/0550-3213(83)90347-4",
    journal = "Nucl. Phys. B",
    volume = "229",
    pages = "528--540",
    year = "1983"
}

@article{Martin:2013tda,
    author = "Martin, Jerome and Ringeval, Christophe and Vennin, Vincent",
    title = "{Encyclop{\ae}dia Inflationaris}: {Opiparous Edition}",
    eprint = "1303.3787",
    archivePrefix = "arXiv",
    primaryClass = "astro-ph.CO",
    doi = "10.1016/j.dark.2024.101653",
    journal = "Phys. Dark Univ.",
    volume = "5-6",
    pages = "75--235",
    year = "2014"
}

@article{Artymowski:2019jlh,
    author = "Artymowski, Michal and Ben-Dayan, Ido",
    title = "{Inflation in supergravity from field redefinitions}",
    eprint = "1908.07052",
    archivePrefix = "arXiv",
    primaryClass = "hep-th",
    doi = "10.3390/sym12050806",
    journal = "Symmetry",
    volume = "12",
    number = "5",
    pages = "806",
    year = "2020"
}

@article{Santos:2022aeb,
    author = "Santos, F. B. M. dos and Silva, R.",
    title = "{Revisiting Witten-O'Raifeartaigh inflation for a non-minimally coupled scalar field}",
    eprint = "2204.13694",
    archivePrefix = "arXiv",
    primaryClass = "astro-ph.CO",
    doi = "10.1088/1475-7516/2022/08/002",
    journal = "JCAP",
    volume = "08",
    number = "08",
    pages = "002",
    year = "2022"
}

@article{Santos:2023hhk,
    author = "Santos, F. B. M. dos and Silva, R. and Alcaniz, J. S.",
    title = "{Constraints on the non-minimally coupled Witten-O'Raifeartaigh inflation}",
    eprint = "2306.07260",
    archivePrefix = "arXiv",
    primaryClass = "astro-ph.CO",
    doi = "10.1088/1475-7516/2023/07/027",
    journal = "JCAP",
    volume = "07",
    pages = "027",
    year = "2023"
}

@article{Drewes:2022nhu,
    author = "Drewes, Marco and Ming, Lei",
    title = "{Connecting Cosmic Inflation to Particle Physics with LiteBIRD, CMB-S4, EUCLID, and SKA}",
    eprint = "2208.07609",
    archivePrefix = "arXiv",
    primaryClass = "hep-ph",
    doi = "10.1103/PhysRevLett.133.031001",
    journal = "Phys. Rev. Lett.",
    volume = "133",
    number = "3",
    pages = "031001",
    year = "2024"
}

@article{Lewis:1999bs,
      author         = "Lewis, Antony and Challinor, Anthony and Lasenby,
                        Anthony",
      title          = "{Efficient computation of CMB anisotropies in closed FRW
                        models}",
      journal        = "\apj",
      volume         = "538",
      year           = "2000",
      pages          = "473-476",
      doi            = "10.1086/309179",
      eprint         = "astro-ph/9911177",
      archivePrefix  = "arXiv",
      primaryClass   = "astro-ph",
      SLACcitation   = "%%CITATION = ASTRO-PH/9911177;%%"
}

@article{Howlett:2012mh,
    author = "Howlett, Cullan and Lewis, Antony and Hall, Alex and Challinor, Anthony",
    title = "{CMB power spectrum parameter degeneracies in the era of precision cosmology}",
    eprint = "1201.3654",
    archivePrefix = "arXiv",
    primaryClass = "astro-ph.CO",
    doi = "10.1088/1475-7516/2012/04/027",
    journal = "JCAP",
    volume = "04",
    pages = "027",
    year = "2012"
}

@software{2019ascl.soft10019T,
       author = {{Torrado}, Jes{\'u}s and {Lewis}, Antony},
        title = "{Cobaya: Bayesian analysis in cosmology}",
 howpublished = {Astrophysics Source Code Library, record ascl:1910.019},
         year = 2019,
        month = oct,
          eid = {ascl:1910.019},
archivePrefix = {ascl},
       eprint = {1910.019},
       adsurl = {https://ui.adsabs.harvard.edu/abs/2019ascl.soft10019T}
}

@article{Torrado:2020dgo,
    author = "Torrado, Jesus and Lewis, Antony",
    title = "{Cobaya: Code for Bayesian Analysis of hierarchical physical models}",
    eprint = "2005.05290",
    archivePrefix = "arXiv",
    primaryClass = "astro-ph.IM",
    reportNumber = "TTK-20-15",
    doi = "10.1088/1475-7516/2021/05/057",
    journal = "JCAP",
    volume = "05",
    pages = "057",
    year = "2021"
}

@article{Planck:2018nkj,
    author = "Aghanim, N. and others",
    collaboration = "Planck",
    title = "{Planck 2018 results. I. Overview and the cosmological legacy of Planck}",
    eprint = "1807.06205",
    archivePrefix = "arXiv",
    primaryClass = "astro-ph.CO",
    doi = "10.1051/0004-6361/201833880",
    journal = "Astron. Astrophys.",
    volume = "641",
    pages = "A1",
    year = "2020"
}

@article{Planck:2018vyg,
    author = "Aghanim, N. and others",
    collaboration = "Planck",
    title = "{Planck 2018 results. VI. Cosmological parameters}",
    eprint = "1807.06209",
    archivePrefix = "arXiv",
    primaryClass = "astro-ph.CO",
    doi = "10.1051/0004-6361/201833910",
    journal = "Astron. Astrophys.",
    volume = "641",
    pages = "A6",
    year = "2020",
    note = "[Erratum: Astron.Astrophys. 652, C4 (2021)]"
}

@article{BICEP:2021xfz,
    author = "Ade, P. A. R. and others",
    collaboration = "BICEP, Keck",
    title = "{Improved Constraints on Primordial Gravitational Waves using Planck, WMAP, and BICEP/Keck Observations through the 2018 Observing Season}",
    eprint = "2110.00483",
    archivePrefix = "arXiv",
    primaryClass = "astro-ph.CO",
    doi = "10.1103/PhysRevLett.127.151301",
    journal = "Phys. Rev. Lett.",
    volume = "127",
    number = "15",
    pages = "151301",
    year = "2021"
}

@article{SimonsObservatory:2018koc,
    author = "Ade, Peter and others",
    collaboration = "Simons Observatory",
    title = "{The Simons Observatory: Science goals and forecasts}",
    eprint = "1808.07445",
    archivePrefix = "arXiv",
    primaryClass = "astro-ph.CO",
    doi = "10.1088/1475-7516/2019/02/056",
    journal = "JCAP",
    volume = "02",
    pages = "056",
    year = "2019"
}

@article{Matsumura:2013aja,
    author = "Matsumura, T. and others",
    title = "{Mission design of LiteBIRD}",
    eprint = "1311.2847",
    archivePrefix = "arXiv",
    primaryClass = "astro-ph.IM",
    doi = "10.1007/s10909-013-0996-1",
    journal = "J. Low Temp. Phys.",
    volume = "176",
    pages = "733",
    year = "2014"
}

@article{LiteBIRD:2022cnt,
    author = "Allys, E. and others",
    collaboration = "LiteBIRD",
    title = "{Probing Cosmic Inflation with the LiteBIRD Cosmic Microwave Background Polarization Survey}",
    eprint = "2202.02773",
    archivePrefix = "arXiv",
    primaryClass = "astro-ph.IM",
    doi = "10.1093/ptep/ptac150",
    journal = "PTEP",
    volume = "2023",
    number = "4",
    pages = "042F01",
    year = "2023"
}

@article{Felder:2000hj,
    author = "Felder, Gary N. and Garcia-Bellido, Juan and Greene, Patrick B. and Kofman, Lev and Linde, Andrei D. and Tkachev, Igor",
    title = "{Dynamics of symmetry breaking and tachyonic preheating}",
    eprint = "hep-ph/0012142",
    archivePrefix = "arXiv",
    reportNumber = "CITA-2000-60, SU-ITP-00-35, IFT-UAM-CSIC-00-40, FT-UAM-00-26, CERN-TH-2000-365",
    doi = "10.1103/PhysRevLett.87.011601",
    journal = "Phys. Rev. Lett.",
    volume = "87",
    pages = "011601",
    year = "2001"
}

@article{Prokopec:1996rr,
    author = "Prokopec, Tomislav and Roos, Thomas G.",
    title = "{Lattice study of classical inflaton decay}",
    eprint = "hep-ph/9610400",
    archivePrefix = "arXiv",
    reportNumber = "CLNS-96-1450, CLNS-96-1438",
    doi = "10.1103/PhysRevD.55.3768",
    journal = "Phys. Rev. D",
    volume = "55",
    pages = "3768--3775",
    year = "1997"
}

@article{Copeland:2002ku,
    author = "Copeland, Edmund J. and Pascoli, S. and Rajantie, A.",
    title = "{Dynamics of tachyonic preheating after hybrid inflation}",
    eprint = "hep-ph/0202031",
    archivePrefix = "arXiv",
    reportNumber = "DAMTP-2002-1, SUSX-TH-02-002, SISSA-2-2002-EP",
    doi = "10.1103/PhysRevD.65.103517",
    journal = "Phys. Rev. D",
    volume = "65",
    pages = "103517",
    year = "2002"
}

@article{Podolsky:2005bw,
    author = "Podolsky, Dmitry I. and Felder, Gary N. and Kofman, Lev and Peloso, Marco",
    title = "{Equation of state and beginning of thermalization after preheating}",
    eprint = "hep-ph/0507096",
    archivePrefix = "arXiv",
    reportNumber = "UMN-TH-2407-05",
    doi = "10.1103/PhysRevD.73.023501",
    journal = "Phys. Rev. D",
    volume = "73",
    pages = "023501",
    year = "2006"
}

@article{Garcia-Bellido:2007fiu,
    author = "Garcia-Bellido, Juan and Figueroa, Daniel G. and Sastre, Alfonso",
    title = "{A Gravitational Wave Background from Reheating after Hybrid Inflation}",
    eprint = "0707.0839",
    archivePrefix = "arXiv",
    primaryClass = "hep-ph",
    reportNumber = "IFT-UAM-CSIC-07-38",
    doi = "10.1103/PhysRevD.77.043517",
    journal = "Phys. Rev. D",
    volume = "77",
    pages = "043517",
    year = "2008"
}

@article{Cuissa:2018oiw,
    author = "Cuissa, Jose Roberto Canivete and Figueroa, Daniel G.",
    title = "{Lattice formulation of axion inflation. Application to preheating}",
    eprint = "1812.03132",
    archivePrefix = "arXiv",
    primaryClass = "astro-ph.CO",
    doi = "10.1088/1475-7516/2019/06/002",
    journal = "JCAP",
    volume = "06",
    pages = "002",
    year = "2019"
}

@article{Kim:2025ikw,
    author = "Kim, Jinsu and Yang, Zihao and Zhang, Ying-li",
    title = "{Gravitational wave signatures of preheating in Higgs-R2 inflation}",
    eprint = "2503.16907",
    archivePrefix = "arXiv",
    primaryClass = "astro-ph.CO",
    doi = "10.1103/pvtn-z4xt",
    journal = "Phys. Rev. D",
    volume = "112",
    number = "4",
    pages = "043534",
    year = "2025"
}

@article{Jia:2024fmo,
    author = "Jia, Tianyu and Sang, Yu and Zhang, Xue",
    title = "{Nonlinear dynamics of oscillons and transients during preheating after single field inflation}",
    eprint = "2409.04046",
    archivePrefix = "arXiv",
    primaryClass = "astro-ph.CO",
    doi = "10.1103/PhysRevD.111.083531",
    journal = "Phys. Rev. D",
    volume = "111",
    number = "8",
    pages = "083531",
    year = "2025"
}

@article{Figueroa:2020rrl,
    author = "Figueroa, Daniel G. and Florio, Adrien and Torrenti, Francisco and Valkenburg, Wessel",
    title = "{The art of simulating the early Universe -- Part I}",
    eprint = "2006.15122",
    archivePrefix = "arXiv",
    primaryClass = "astro-ph.CO",
    doi = "10.1088/1475-7516/2021/04/035",
    journal = "JCAP",
    volume = "04",
    pages = "035",
    year = "2021"
}

@article{Figueroa:2021yhd,
    author = "Figueroa, Daniel G. and Florio, Adrien and Torrenti, Francisco and Valkenburg, Wessel",
    title = "{CosmoLattice: A modern code for lattice simulations of scalar and gauge field dynamics in an expanding universe}",
    eprint = "2102.01031",
    archivePrefix = "arXiv",
    primaryClass = "astro-ph.CO",
    doi = "10.1016/j.cpc.2022.108586",
    journal = "Comput. Phys. Commun.",
    volume = "283",
    pages = "108586",
    year = "2023"
}

@article{Figueroa:2023xmq,
    author = "Figueroa, Daniel G. and Florio, Adrien and Torrenti, Francisco",
    title = "{Present and future of ${\mathcal{C}}$ osmo ${\mathcal{L}}$ attice}",
    eprint = "2312.15056",
    archivePrefix = "arXiv",
    primaryClass = "astro-ph.CO",
    doi = "10.1088/1361-6633/ad616a",
    journal = "Rept. Prog. Phys.",
    volume = "87",
    number = "9",
    pages = "094901",
    year = "2024"
}

@article{Dux:2022kuk,
    author = "Dux, Fr{\'e}d{\'e}ric and Florio, Adrien and Klari{\'c}, Juraj and Shkerin, Andrey and Timiryasov, Inar",
    title = "{Preheating in Palatini Higgs inflation on the lattice}",
    eprint = "2203.13286",
    archivePrefix = "arXiv",
    primaryClass = "hep-ph",
    reportNumber = "FTPI-MINN-22-08, UMN-TH-4117/22, CP3-22-28",
    doi = "10.1088/1475-7516/2022/09/015",
    journal = "JCAP",
    volume = "09",
    pages = "015",
    year = "2022"
}

@article{Schmitz:2020syl,
    author = "Schmitz, Kai",
    title = "{New Sensitivity Curves for Gravitational-Wave Signals from Cosmological Phase Transitions}",
    eprint = "2002.04615",
    archivePrefix = "arXiv",
    primaryClass = "hep-ph",
    reportNumber = "CERN-TH-2020-018",
    doi = "10.1007/JHEP01(2021)097",
    journal = "JHEP",
    volume = "01",
    pages = "097",
    year = "2021"
}

@article{Antusch:2025ewc,
    author = "Antusch, Stefan and Marschall, Kenneth and Torrenti, Francisco",
    title = "{Equation of state during (p)reheating with trilinear interactions}",
    eprint = "2507.13465",
    archivePrefix = "arXiv",
    primaryClass = "astro-ph.CO",
    doi = "10.1088/1475-7516/2025/11/002",
    journal = "JCAP",
    volume = "11",
    pages = "002",
    year = "2025"
}

@article{Ueno:2016dim,
    author = "Ueno, Yoshiki and Yamamoto, Kazuhiro",
    title = "{Constraints on $\alpha$-attractor inflation and reheating}",
    eprint = "1602.07427",
    archivePrefix = "arXiv",
    primaryClass = "astro-ph.CO",
    reportNumber = "HUPD-1601",
    doi = "10.1103/PhysRevD.93.083524",
    journal = "Phys. Rev. D",
    volume = "93",
    number = "8",
    pages = "083524",
    year = "2016"
}

@article{Drewes:2017fmn,
    author = "Drewes, Marco and Kang, Jin U and Mun, Ui Ri",
    title = "{CMB constraints on the inflaton couplings and reheating temperature in $\alpha$-attractor inflation}",
    eprint = "1708.01197",
    archivePrefix = "arXiv",
    primaryClass = "astro-ph.CO",
    reportNumber = "TUM-HEP-1093-17",
    doi = "10.1007/JHEP11(2017)072",
    journal = "JHEP",
    volume = "11",
    pages = "072",
    year = "2017"
}

@article{Drewes:2023bbs,
    author = "Drewes, Marco and Ming, Lei and Oldengott, Isabel",
    title = "{LiteBIRD and CMB-S4 sensitivities to reheating in plateau models of inflation}",
    eprint = "2303.13503",
    archivePrefix = "arXiv",
    primaryClass = "hep-ph",
    doi = "10.1088/1475-7516/2024/05/081",
    journal = "JCAP",
    volume = "05",
    pages = "081",
    year = "2024"
}

@article{Das:2025teu,
    author = "Das, Suratna and Kumar, Umang and Mishra, Swagat S. and Sahni, Varun",
    title = "{Witten-O{\textquoteright}Raifeartaigh potential revisited in the context of warm inflation}",
    eprint = "2511.23219",
    archivePrefix = "arXiv",
    primaryClass = "astro-ph.CO",
    doi = "10.1103/mjx4-kgns",
    journal = "Phys. Rev. D",
    volume = "113",
    number = "6",
    pages = "063522",
    year = "2026"
}

\end{document}